\documentclass[fleqn,usenatbib]{mnras}

\usepackage{newtxtext,newtxmath}
\usepackage{subfigure}
\usepackage{multirow}
\usepackage{natbib}
\usepackage{enumerate}
\usepackage[normalem]{ulem}

\usepackage[T1]{fontenc}
\usepackage{float}

\DeclareRobustCommand{\VAN}[3]{#2}
\let\VANthebibliography\thebibliography
\def\thebibliography{\DeclareRobustCommand{\VAN}[3]{##3}\VANthebibliography}

\usepackage{graphicx}	
\usepackage{amsmath}	
\usepackage{comment}

\title[ML for accreted vs in-situ classification]{Applying machine learning to Galactic Archaeology: how well can we recover the origin of stars in Milky Way-like galaxies?}

\author[A. Sante et al.]{
Andrea Sante,$^{1}$\thanks{E-mail: A.Sante@2022.ljmu.ac.uk}
Andreea S. Font,$^{1,2}$
Sandra Ortega-Martorell,$^{2}$
Ivan Olier$^{2}$, Ian G. McCarthy$^{1,2}$
\\
$^{1}$Astrophysics Research Institute, Liverpool John Moores University, 146 Brownlow Hill, Liverpool L3 5RF, UK\\
$^{2}$Data Science Research Centre, Liverpool John Moores University, 3 Byrom Street, Liverpool L3 3AF, UK
}

\date{Accepted XXX. Received YYY; in original form ZZZ}

\pubyear{2023}

\begin{document}
\label{firstpage}
\pagerange{\pageref{firstpage}--\pageref{lastpage}}
\maketitle

\begin{abstract}
We present several machine learning (ML) models developed to efficiently separate stars formed in-situ in Milky Way-type galaxies from those that were formed externally and later accreted. These models, which include examples from artificial neural networks, decision trees and dimensionality reduction techniques, are trained  on a sample of disc-like, Milky Way-mass galaxies drawn from the \texttt{ARTEMIS} cosmological hydrodynamical zoom-in simulations.  We find that the input parameters which provide an optimal performance for these models consist of a combination of stellar positions, kinematics, chemical abundances ([Fe/H] and [$\alpha$/Fe]) and photometric properties.  Models from all categories perform similarly well, with area under the precision-recall curve (PR-AUC) scores of $\simeq 0.6$. Beyond a galactocentric radius of $5$~kpc, models retrieve  $>90\%$ of accreted stars, with a sample purity close to $60\%$, however the purity can be increased by adjusting the classification threshold. For one model, we also include host galaxy-specific properties in the training, to account for the variability of accretion histories of the hosts, however this does not lead to an improvement in performance. The ML models can identify accreted stars even in regions heavily dominated by the in-situ component (e.g., in the disc), and perform well on an unseen suite of simulations (the \texttt{Auriga} simulations). The general applicability bodes well for application of such methods on observational data to identify accreted substructures in the Milky Way  without the need to resort to selection cuts for minimising the contamination from in-situ stars. 
\end{abstract}




\section{Introduction}

In a $\Lambda$ cold dark matter ($\Lambda$CDM) cosmological model, large galaxies like the Milky Way form through a hierarchical process, with smaller structures merging progressively into larger ones \citep{white_core_1978,searle_composition_1978}. In this framework, signatures of past accretion and disruption events in the Galaxy are left imprinted in a multi-dimensional parameter space, composed of positions, kinematics and chemical abundances of stars (e.g., \citealt{Helmi2020}). From the information gathered about tidal streams in this multi-dimensional parameter space, one can reconstruct the assembly history of the Milky Way \citep{Freeman2002}, i.e., determine the timing of the accretion events, the masses of the progenitor galaxies, their star formation histories or the orbital properties. 

A multitude of methods have been devised to find tidal stellar streams. Streams from massive accretions can be usually detected from photometry, as they tend to be brighter and to be spatially extended. Those from low mass progenitors may sometimes appear as coherent structures in physical space (i.e., in positions and velocities of stars), particularly if the accretions were recent (e.g., \citealt{bullock_tracing_2005,Johnston2008}) or in a special geometry \citep{Johnston1996}. Over longer dynamical timescales, however, the streams tend to disperse due to phase-mixing \citep{tremaine_geometry_1999} and thus they become increasingly difficult to distinguish from background field stars. Information about them may be still retained in the integrals of motion related to angular momenta and total energies of their orbits (see \citealt{Binney2008}). Methods have been devised to identify tidal streams as "clumps" in the energy ($E$) and angular momentum ($L_z$) space \citep{helmi_building_1999,Gomez2010}, as these quantities are (quasi-)conserved through time. Methods that rely on finding specific patterns of tidal streams in the velocity space \citep{Johnston2002,Gomez2010,Koppelman2021}, or in the angles, actions or frequencies  \citep{Bovy2014,McMillan2008,Sanders2016,malhan_2022} have also been used to identify streams in the Galaxy. However, many of these methods require knowledge  of the gravitational potential of the Milky Way, although more recent techniques, such as \texttt{STREAMFINDER} \citep{malhan_streamfinder_2018}, do not rely on such assumptions. 

Other parameters have also been used to improve the detection. For example, methods based on match-filters that weigh the colour-magnitudes of stars \citep{Grillmair1995,Rockosi2002,Balbinot2011} have proven useful in the detection of new streams \citep{Shipp2019}.  More widely used are methods employing the chemical abundances of stars, building on the expectation that stars formed in a given progenitor share similar chemical ``fingerprints'' even if their spatial and kinematical information has become phase-mixed. In combination with 6D physical space parameters (positions and kinematics), the principle of ``chemical tagging'' \citep{Freeman2002} has been applied successfully in the discovery or characterisation of tidal debris \citep{helmi_merger_2018,belokurov_co-formation_2018,Koppelman2019,Kruijssen2019,Das2020,Ji2020,Naidu2022,horta_chemical_2023,malhan_shiva_shakti_2024}, or in the reconstruction of the early stages of the Milky Way \citep{belokurov_-situ_2023}.

In the era of large Galactic surveys, such as {\it Gaia},  new methods based on machine learning (ML) techniques have been developed and have proven viable. Clustering algorithms, such as DBSCAN \citep{Ester1996} have been applied to chemo-dynamical data to confirm existing discoveries or to reveal new ones  \citep{Koppelman2019,borsato_identifying_2020}. An unsupervised learning method, called \texttt{VIA MACHINAE}, was also developed, using conditional density estimation and sideband interpolation to find local overdensities \citep{shih_via_2022}.  
\citet{veljanoski_leaves_2019} developed a gradient boosted trees model to identify halo stars based on astrometric and photometric data from the \texttt{Gaia Universe Model Snapshot} \citep{robin_gums_2012}. 

The discovery of new tidal streams is becoming increasingly more difficult, in part due to current limitations in the current methods. For example, many stream-finding methods assume that the halo is mostly of accreted origin. This a reasonable assumption only for the outer parts of the Galaxy ($\gtrapprox 20$~kpc), which are less contaminated by disc stars. Observations indicate that the stellar halo has a ``dual nature'' \citep{Carollo2007,Carollo2010,Beers2012}, where the two components, accreted and in-situ, overlap over some distance. Disentangling the two components is important in order to construct a relatively clean sample of accreted stars on which to apply the stream detection methods. The two components differ in spatial distribution, kinematics and metallicity \citep{Carollo2007}; specifically, the in-situ halo is more centrally concentrated than the accreted and tends to have an overall rotating motion prograde with the disc, whereas the orbits of accreted stars are more randomly distributed; generally, the in-situ halo is also more metal-rich than the accreted. This suggests that the two halo components could be, in principle, clearly separated. In practice, however, most observational samples include some selection criteria which are meant to minimise the contamination from both the disc and in the situ halo.  

The observational selection cuts usually relate to spatial location, kinematics or chemistry (or a combination thereof). In some cases, the criteria are purposefully conservative, for example, selecting stars only on retrograde orbits to search for debris. This kinematical cut has proven beneficial for the discovery of many substructures in the solar neighbourhood \citep{Koppelman2019}, including a debris from a massive progenitor, called Gaia Enceladus/Sausage (GES) \citep{helmi_merger_2018,belokurov_co-formation_2018}, which merged with the Galaxy $\gtrapprox 8-9$~Gyr ago. However, many tidal streams are predicted to still remain hidden  \citep{Shipp2023}. Therefore, relaxing the selection criteria for observational samples could lead to more debris discoveries, especially in the less explored regions of the Galaxy, such as the heated stellar disc \citep{mackereth_dynamical_2019,belokurov_biggest_2020}.

The dual nature of stellar haloes is retrieved naturally in cosmological hydrodynamical simulations \citep{Zolotov2009,mccarthy_global_2012,Tissera2013,Cooper2015,Pillepich2015,Monachesi2016,Pillepich2018,Brook2020}. However, these simulations are too general to inform the precise selection cuts that can be applied to observations in the Milky Way, as they model systems with a variety of different accretion histories none of which is expected to exactly match that of the Milky Way. Moreover, depending on the implementation of sub-grid physical prescriptions, simulations may predict different properties for the in-situ halo component. This is related to the different formation channels of in-situ halo stars in simulations: either as stars ejected from the galaxy disc, or formed within filaments of cold gas, or in the wakes of stripped gas from infalling satellites. 

Rather than using simulations to inform selection cuts, one can use them to train ML models to separate the accreted from in-situ stars more accurately and hopefully in a way which is sufficiently general to apply to the observational data. ML provides an ideal framework to find out the relations between objects belonging to different classes by leveraging the information hidden in large datasets. This is particularly useful in those regions of the parameter space where the two halo components overlap, for example, in the case of accreted stars that overlap with the disc (e.g., \citealt{Hawkins2015}), or of the old, in-situ halo stars that may overlap in metallicity with some of the accreted substructure. 

Several ML techniques have been developed recently to separate the two components. For example, by analysing a {\it Gaia} mock catalog constructed from the \texttt{FIRE} simulations \citep{sanderson_synthetic_2020}, \citet{ostdiek_cataloging_2020} trained an artificial neural network (ANN) to classify accreted and in-situ stars based on 5D kinematics and then fine-tuning the model on a {\it Gaia} DR2/ RAVE data set. This has led to the discovery of a new substructure in the Milky Way, called Nyx \citep{necib_chasing_2020}. Recently, \citet{tronrud_machine_2022} developed an ANN to separate accreted and in-situ stars, and trained it on the chemical abundances and ages of stars in \texttt{Auriga} simulations \citep{grand_auriga_2017}. A similar method has been developed by \citet{trujillo_gomez_2023} to classify accreted and in-situ globular clusters in the \texttt{E-MOSAICS} simulations \citep{Pfeffer2018}, using as inputs 17 observable properties, including some of the associated host galaxies. 

Rather than focusing on the description of a single methodology, here we perform an extensive comparison of different ML algorithms, in order to decide which ones are more suitable for the classification of accreted and in-situ stars. We also include a wide range of input parameters the models, chosen as stellar parameters which can be directly observed from Milky Way surveys, such as positions, kinematics, ages, chemical abundances ([Fe/H] and [$\alpha$/Fe]) and photometric properties. Our aim is to identify the {\it optimal, data-driven model} that can automatically identify accreted stars in observational samples of the Milky Way. The ML algorithms we consider can be grouped broadly into three categories: ANNs, decision trees, and dimensionality reduction methods. The ANNs and decision trees are used for developing the classification models, whereas the dimensionality reduction technique is used for visualising the data and providing insights into the output of the models. As a benchmark for comparison of models, we adopt an ANN model which resembles the ``Galactic Archaeology Neural Network'' of \citet{tronrud_machine_2022}. For training and testing the ML models, we use a sample Milky Way-mass galaxies from the \texttt{ARTEMIS} simulations \citep{font_artemis_2020}, selected to be disc-like today, but with different accretion histories. Since in simulations the origin of each star particle is already known (i.e., whether it formed in-situ or was accreted), we can assess the performance of the techniques more accurately than by testing them on observations. 

The paper is organised as follows. Section \ref{sec:data} provides a brief description of the \texttt{ARTEMIS} simulations and of the sample of disc-like galaxies which are used for training and testing. The selection of the physical parameters considered as inputs for the models is described in Section~\ref{sec:feature_selection} and Section~\ref{sec:galaxy_feature_selection}. In Section \ref{sec:ML} we present the ML methods; the metrics used for evaluating the classification are discussed in Section~\ref{subsec:metrics}, while the methods are described individually in Section ~\ref{subsec:models}, including a description of how we determine the optimal set of input parameters (in Section~\ref{subsec:optimal_feat}). Section~\ref{sec:results} includes a comparison of the performance of the ML models (Section~\ref{subsec:comparison_ML}) and shows how the in-situ and accreted stars identified by these methods are separated in a chemo-dynamical phase-space (Section~\ref{subsec:fp_fn}); it also shows how ML models may improve the detection of accreted stars in areas omitted by observational selection cuts  (Section~\ref{subsec:selection_cuts}); and illustrates how the predictions of the models can be visualized with UMAP (Section~\ref{subsec:umap_results}). In Section~\ref{sec:cross_sim} we apply our ML methods on a different suite of simulations (\texttt{Auriga}), to further test their performance. The conclusions of our study are summarized in Section~\ref{sec:conclusions}.

\section{The ARTEMIS Simulations}
\label{sec:data}

\texttt{ARTEMIS} is a suite of zoomed-in, high-resolution cosmological hydrodynamical simulations of $45$ Milky Way-mass systems \citep{font_artemis_2020,Font2021} in a flat $\Lambda$CDM WMAP9 \citep{hinshaw2013} cosmological model, with the following parameters: $\Omega_\textrm{m}=0.2793$, $\Omega_\textrm{b}=0.0463$, $h=0.70$, $\sigma_8=0.8211$ and $n_s=0.972$. The Milky Way-mass systems have total masses ranging between $8 \times 10^{11} < {M}_{200}/{\rm M}_{\sun} < 2 \times 10^{12}$, where ${M}_{200}$ is the mass enclosing a mean density of $200$ times the critical density of the Universe at present time. The dark matter particle masses are $1.17\times10^5$ M$_{\sun}\/h^{-1}$, the initial gas particle masses are $2.23\times10^4$ M$_{\sun}\/h^{-1}$, and the (Plummer equivalent) force resolution is $125$~pc $h^{-1}$. 

The simulations were run with the Gadget-3 code \citep{Springel2005}, including an updated hydrodynamical solver and subgrid physical prescriptions developed for the \texttt{EAGLE} project \citep{Schaye2015}. The physical prescriptions for subgrid physics include metal-dependent radiative cooling in the presence of a photo-ionizing UV background, star formation, stellar and chemical evolution, formation of supermassive black holes, and feedback from supernova, stellar winds and active galactic nuclei (see \citealt{Crain2015} and \citealt{Schaye2015} for details).
The chemical enrichment model follows 11 element species taking in account the mass loss from AGB stars, stellar winds and both core collapse and thermonuclear supernovae. The heavy elements relevant for our study, are Fe (produced mainly in Type Ia SNe) and Mg (an $\alpha$-element, produced in Type II SNe).

\citet{font_artemis_2020} also computed, in post-processing, the optical properties of the simulated galaxies in \texttt{ARTEMIS}. These include luminosities, magnitudes and colours in various passbands, assuming each star particle is an a single stellar population, by using the \texttt{PARSEC v1.2S + COLIBRI PR16} isochrones \citep{Bressan2012,Marigo2017} and a Chabrier \citep{Chabrier2003} initial mass function. In this study we are mainly interested in \textit{Gaia} photometric properties, which were not computed originally. For this, we use the available SDSS magnitudes and convert them to the \textit{Gaia} $G$, $G_{\mathrm{BP}}$, and $G_{\mathrm{RP}}$ passbands equivalents, using the photometric relationships from \citet{gaiaDR3_doc}.

The origin of each star particle (accreted/ in-situ) was determined in post-processing. Here we use the labels from \citet{font_artemis_2020}.  In brief, the redshift of formation is recorded for each star particle during the simulation and in post processing the simulation snapshot that this is closest to (in lookback time) was identified.  If at the time of its formation the star particle was gravitationally bound to the main progenitor of the Milky Way-mass galaxy, it was labeled as in-situ, otherwise, as accreted. Note that by this definition, stars that are born from gas stripped from an infalling satellite, but inside the main halo, are also labeled in-situ. This is the case only for a small percentage of stars, however.

Since we aim to devise ML models suited for the Galactic observations, we focus our training and testing of models on a subset of galaxies from  \texttt{ARTEMIS} which have a disc-like component similar to that of the Milky Way\footnote{We note that, in an initial phase, we used for training simulated galaxies irrespective of their morphological type (i.e., both ellipticals and discs). However, we found that the performance of models was significantly reduced in this case, particularly when the training set was composed of mostly different morphological types than the testing set. Since here we are mainly interested in optimising the performance of the ML models, we choose to use a sub-sample composed of only disc-like systems. This restricts somewhat the accretion histories that are included in the datasets to those that are more quiescent. However, this approach is justified as: 1) the accretion history of the Milky Way is partially known, and therefore we do not need to be completely agnostic about this feature; 2) disc-like galaxies can form with a variety of accretion histories (see, e.g., \citealt{Font2017}) and our disc-like subsample covers a broad range of these scenarios.}. This ensures that there are sufficient examples in the learning set which encapsulate the distribution of accreted and in-situ stars in a disc-like galaxy. For the disc-like selection criteria, we use the kinematics, specifically the co-rotational parameter $\kappa_{\rm co}$ (\citealt{abadi_simulations_2003}; see also \citealt{font_artemis_2020} and \citealt{dillamore_merger-induced_2022}). Here we define it as $\kappa_{\rm co}=  \sum_{r=0}^{30\mathrm{kpc}} L_{z} \, / \, \sum_{r=0}^{30\mathrm{kpc}} L_{z,circ}$, where $L_{z}$ is the total stellar angular momentum along the $z$-axis, and  $L_{z,circ}$ is the total angular momentum of star particles with the same energy but in a co-rotating circular orbit. For this computation we only consider star particles within an aperture of $30 \, \mathrm{kpc}$, and impose a cut-off of $\kappa_{\rm co} \geq 0.50$ to select galaxies with the most prominent disc components. This results in a sample of $16$ galaxies listed in Table \ref{tab:datasets}, together with their main physical properties: the total accreted stellar fraction, the co-rotation parameter, the total stellar mass, half -(stellar) mass radius, maximum circular velocity, and average chemical abundances ([Fe/H] and [$\alpha$/Fe]).

\begin{table*}
\begin{center}
\begin{tabular}[H]{| c c c c c c c c |}
    \hline
    Galaxy & $f_{acc}$ & $k_{\mathrm{\rm co}}$ & $M_{*} [10^{10} \, \mathrm{M}_{\odot}]$ & $r_{1/2} \, [\mathrm{kpc}]$ & $v_{\theta_{\rm MAX}} [\mathrm{km} \mathrm{s}^{-1}]$ & $\langle [\mathrm{Fe}/\mathrm{H}] \rangle$ & $\langle [\mathrm{\alpha}/\mathrm{Fe}] \rangle$\\
    \hline
    \multicolumn{8}{| c |}{Training dataset}\\
    \hline
    G01 & 0.14 & 0.60 & 3.64 & 4.86 & 199 & -0.14 & 0.19 \\
 
    G15 & 0.11 & 0.61 & 3.57 & 5.88 & 170 & -0.16 & 0.22 \\

    G17 & 0.10 & 0.69 & 3.74 & 7.26 & 198 & -0.21 & 0.25 \\

    G18 & 0.22 & 0.59 & 2.78 & 4.38 & 184 & -0.15 & 0.20 \\

    G19 & 0.04 & 0.67 & 2.57 & 4.92 & 177 & -0.16 & 0.24 \\

    G23 & 0.11 & 0.56 & 2.87 & 2.77 & 197 & -0.07 & 0.21 \\

    G24 & 0.11 & 0.55 & 3.63 & 3.90 & 185 & -0.16 & 0.22 \\

    G25 & 0.18 & 0.63 & 2.57 & 5.52 & 172 & -0.24 & 0.26 \\

    G27 & 0.21 & 0.57 & 2.57 & 5.40 & 160 & -0.19 & 0.22 \\

    G38 & 0.04 & 0.81 & 2.97 & 8.46 & 176 & -0.13 & 0.20 \\

    G40 & 0.16 & 0.64 & 2.02 & 4.50 & 155 & -0.15 & 0.20 \\

    G44 & 0.12 & 0.63 & 4.28 & 5.22 & 204 & -0.27 & 0.31 \\
    \hline
    \multicolumn{8}{| c |}{Test dataset}\\
    \hline
    G29 & 0.08 & 0.65 & 2.95 & 2.60 & 210 & -0.08 & 0.18 \\

    G30 & 0.28 & 0.55 & 2.12 & 4.20 & 172 & -0.10 & 0.23 \\

    G34 & 0.05 & 0.78 & 2.76 & 6.20 & 183 & -0.16 & 0.21 \\

    G42 & 0.13 & 0.65 & 2.10 & 3.10 & 174 & -0.18 & 0.25 \\
    \hline 
\end{tabular} 
\end{center}
    \caption{Sample of disc-like galaxies in \texttt{ARTEMIS} selected based on their co-rotation parameter, $\kappa_{\rm co}$. These galaxies are separated into two datasets used for training and test the performance of the ML models, respectively. The columns are: 1) galaxy label; 2) fraction of accreted stellar component (defined as the mass fraction of accreted star particles over the total stellar mass (in-situ + accreted); 3) co-rotational parameter; 4) the total stellar mass; 5) half stellar mass radius (defined as the radius enclosing 50\% of the total stellar mass); 6) maximum circular velocity; 7) average [Fe/H] abundance; 8) average [$\alpha$/Fe] abundance, where $\alpha$ is tracked by Mg abundance. Apart for the fraction of accreted stars, all quantities are computed within $30 \, \mathrm{kpc}$ from the centre of the MW-mass galaxy.}
    \label{tab:datasets}
\end{table*}

Galaxies are further split into two sets: a training and a test dataset, respectively. The training set is used to provide examples of accreted and in-situ stars to the ML models, while the test dataset is used to assess the classification performance. The test dataset is composed of galaxies with an assembly history more similar (although not exactly the same) to the one inferred for the Milky Way, specifically those where the most massive accreted progenitor (MMAP) was accreted more than $8 \, \textrm{Gyr}$ ago, and the stellar mass ratio of this MMAP to the total stellar mass of the host is $\geq 0.4$ (see figure 7 and Table A1 of \citealt{dillamore_merger-induced_2022} for details). The two datasets, comprising of $12$ galaxies for training and $4$ for testing, are listed separately in Table~\ref{tab:datasets}.

For each galaxy in the training dataset, we also reserve 20\% of stars\footnote{Although our ML models are developed with the aim of applying them on individual stars in the Galaxy, the simulations can only track star particles, which are essentially individual single stellar populations (SSP). For brevity, throughout the paper, we will refer to the star particles as `stars'.} for the validation dataset. This comprises data that are used during the training routine for evaluating the neural networks on unseen data and detecting overfitting, i.e., the modelling of the noise contained in the training examples. 

\subsection{Stellar parameters as features}
\label{sec:feature_selection}

For training and testing of the ML models, each star particle is described as a vector of physical parameters (also known as features) which are expected to be relevant for the distinction between accreted and in-situ stars. Although the simulations provide more information on each star particle (including, e.g., the mass or the gravitational potential), we focus only on stellar parameters which can be observed, to facilitate future applications on survey data. These features are divided into four categories, resembling (very broadly) the focus of different types of observational Galactic surveys:

\begin{itemize}

\item {\it Positions and kinematics.} For many Milky Way stars, positions and velocities are readily available, e.g., from {\it Gaia} and RAVE. Accreted stars are expected to differ from in-situ ones both in terms of their locations and of their overall motions. Accreted stars extend much further out into the halo where they tend to appear as kinematically cold tidal streams, and their orbits tend to be more randomly distributed, whereas the in-situ stars are more prominent in the inner region of a galaxy \citep{font_cosmological_2011}, follow a more flattened distribution, and tend to have a prograde rotation with the disc \citep{mccarthy_global_2012}. Therefore, for positions, we choose as features the Galactocentric radius in the plane of the disc, $R$,  and the distance perpendicular to the plane of the disc, $z$, while for kinematics measures we use the rotational velocity in the plane of the disc, $v_{\theta}$, and the velocity dispersion in the plane perpendicular to the disc, $\sigma$. 

\item {\it Chemical abundances.} Chemical abundances are related to the star formation history of the associated progenitors \citep{Freeman2002}. Here we focus on the stellar metallicities, defined as [Fe/H], and on the [$\alpha$/Fe] abundances, where $\alpha$ is tracked by Mg. We expect stars of accreted origin to have, on average, lower [Fe/H] and higher [Mg/Fe] values than those in-situ.

\item{Stellar ages.}  Stellar ages, $\tau$, can also be used in tandem with the kinematic and chemical properties of stars to trace populations formed in the same galactic environment \citep{Helmi2020}. [Fe/H] values are expected to correlate well with ages and they are often used as proxies for the latter. Old accreted stars are also expected to have higher [$\alpha$/Fe] abundances due to the short, bursty star formation episodes in their parent dwarf galaxies at high redshift \citep{Robertson2005}. In addition, ages can provide complementary information about the rate of chemical enrichment at different epochs (e.g., \citealt{Hawkins2014}).

\item {\it Photometry.} Tidal streams from different disrupted satellite galaxies are expected to stand out in terms of their surface brightness; specifically, the brightness of a stream tends to correlate with the stellar mass of its dwarf progenitor \citep{Font2006b,Johnston2008,Gilbert2009,Cooper2010}. Similarly, accreted stars can also be distinguished from in-situ stars in terms of their photometric properties. Accreted debris tends to be fainter than the in-situ component, due to its lower mass and larger spatial extent. For photometric properties, we choose the absolute magnitude in the {\it Gaia} $\mathrm{G}$ passband, $\mathrm{M}_{\mathrm{G}}$, and the colour evaluated in the $G_{\mathrm{BP}}$ and $G_{\mathrm{RP}}$ passbands, ${\mathrm{BP}}-{\mathrm{RP}}$.  We note, however, that our results are not dependent to the specific {\it Gaia} passbands or survey. 
These photometric properties also correlate with stellar ages (which be inferred from colour-magnitude diagram fitting, e.g., \citealt{Gallart2005}), thus offering an alternative to direct age measurements, which are usually more difficult to obtain.
\end{itemize}

To summarise, we choose as possible input features for ML models the following stellar parameters:

\begin{equation*}
\{R, \, z, \, v_{\theta}, \, \sigma, \mathrm{[Fe/H]}, \, [\alpha/\mathrm{Fe}], \, \tau,  \, \mathrm{M}_{\mathrm{G}}, \, {\mathrm{BP}}-{\mathrm{RP}} \}.   
\end{equation*}

As described later, in Section~\ref{subsec:optimal_feat}, the ML models ultimately include an {\it optimal} set of features, which provides the best performance for our benchmark model. The optimal set is the same as the set of parameters above, but excluding the stellar ages ($\tau$), for which the benchmark model is able to retrieve the information from the other parameters, mainly from [Fe/H] and [$\alpha$/Fe].

\subsection{Galaxy-specific features}
\label{sec:galaxy_feature_selection}

In addition to stellar features, one of our models (see \ref{sec:MLP_galaxyfeatures})
 includes a set of galaxy-specific features, devised to account for the accretion histories of MW-mass hosts. These are listed in Table~\ref{tab:datasets}, and comprise of: the stellar mass half-radius, the co-rotational parameter, the total stellar mass, the maximum circular velocity, and the average [Fe/H] and [$\alpha$/Fe] abundances of the respective MW-mass galaxy. 

In general, we expect galaxies that experienced more massive mergers to have less well-defined stellar discs (e.g., smaller sizes, lower $\kappa_{\rm co}$) and also tend to be more massive. Additionally, we expect that systems with higher masses would be more more chemically enriched (higher $\langle$[Fe/H]$\rangle$ and lower $\langle$[$\alpha$/Fe]$\rangle$). Therefore, some indication about the accretion histories can be inferred from the present-day properties of stellar populations in the MW-mass hosts (see also 
\citealt{Grimozzi2024}).

We note that, although the merger histories of the simulated galaxies are already known (e.g., in the form of merger trees , or the properties of the MMAPs), we choose not to use this type of information and focus instead on observable parameters, as we do in the case of star particle features. In Section~\ref{subsec:comparison_ML} we discuss the performance of models with and without these galaxy-specific features.

\section{Machine learning models}
\label{sec:ML}

To devise an appropriate ML model for our task, we need to address two challenges: {\it i)} to identify ML models which can learn effectively the underlying patterns in the data; and {\it ii)} to determine the input parameters that optimize the performance of these models. 

For the first task, we consider models representative of two main families of supervised ML methods, namely Artificial Neural Networks (ANNs) and decision trees. From the ANN type, we consider a feedforward model called a multilayer perceptron (MLP). Given the relatively small number of features that describe the accreted and in-situ stars, we choose to consider shallower architectures than the ones selected in \citet{tronrud_machine_2022}, so as to limit the risk of overfitting. We then augment this MLP model with domain inputs, i.e., we include additional galaxy-specific features. With the extended set of features, both stellar and galaxy-specific, we aim to mitigate the potential decrease in the performance caused by the variability introduced by the specific assembly histories of galaxies. For the same purpose, we also consider the transformational machine learning (TML) technique \citep{olier_transformational_2021}. From the category of decision tree-like systems, we consider the eXtreme Gradient Boosting (XGBoost) model \citep{chen_xgboost_2016}. 

To better visualise the dataset used for training the models, as well as to understand the functioning of these models, we use the Uniform Manifold Approximation and Projection (UMAP) dimensionality reduction technique \citep{mcinnes_umap_2018}. This method consists of mapping the accreted and in-situ stars into a new, lower-dimensional plane while maintaining the global and local structures; thus, stars are clustered in structures which make relations hidden in physical space more visually evident.

ANNs have been investigated recently for this task \citep{ostdiek_cataloging_2020,tronrud_machine_2022,trujillo_gomez_2023}, and are commonly used in astrophysics, e.g., for the classification of transients and variable stars \citep{jayasinghe_asas-sn_2019, agarwal_fetch_2020, chen_uncloaking_2021}, of quasars \citep{nakoneczny_catalog_2019, clarke_identifying_2020, nakoneczny_photometric_2021}, or of galaxies \citep{Traven2017, dominguez_sanchez_improving_2018, huertas-company_dawes_2023}. XGBoost is also commonly used, with many applications in Galactic studies (e.g., \citealt{anders2023}). To our knowledge, models like TML or UMAP are investigated for the first time here for an astrophysical problem.

For the task of determining the \textit{optimal} set of features, we start with a wide range of physical stellar parameters (described in Section~\ref{sec:feature_selection}) and determine which combination provides the best performance (Section~\ref{subsec:optimal_feat}) for our benchmark model (Section~\ref{subsec:benchmark}).   
The implicit assumption in our approach is that the optimal set of features would be the same for any type of ML model adopted. We then compare the classification performances of different models using the same (fixed) set of features. 

In the following, we describe the metrics used for evaluating the classification performance of models (Section~\ref{subsec:metrics}). The model implementations are described separately in Section \ref{subsec:models}, where we also provide some technical background on each of them. Otherwise specified, we will make use of common ML terminology. 

The training and implementation of the ANNs is performed using the \texttt{TensorFlow} \citep{abadi_tensorflow} library. The XGBoost and UMAP methods are developed using the \texttt{xgboost} \citep{chen_xgboost_2016} and \texttt{umap} \citep{mcinnes2018umap-software} python packages, respectively. 

\subsection{Performance metrics}
\label{subsec:metrics}

In the development of all models we adopt a supervised learning approach. In supervised models, the mapping between features and prediction is learned by providing a set of example-label pairs. Furthermore, the model parameters are tuned to minimise the difference between the prediction and the actual class (the label) as quantified by an objective function. The separation of the accreted versus in-situ stars can be thought as a binary classification problem, where the positive class is represented by the accreted stars and the negative class by the in-situ. A ML model achieves this by applying a sequence of mathematical operations and tunable parameters to map the properties of stars to a value of either $0$ or $1$, representing the negative and positive classes, respectively. 

For the ANN and decision tree models, the prediction for a given star is represented by the output, which is a value between 0 and 1 and measures the probability of the star belonging to the positive (accreted) class. In both cases, a star is classified as accreted if its prediction is greater than a threshold value. The optimal performance of a classifier may occur at a different threshold value for different models, especially in problems with highly imbalanced datasets, such as ours where accreted stars comprise, on average, ~10\% of the overall stellar content of a galaxy. We therefore explore also the effect of changing threshold values on the performance of our models (see Section~\ref{subsec:fp_fn}). 

Here use the usual performance metrics, namely the precision ($P$) and recall ($R$) of a model. These are used to compare the performances of various models, but also to identify the optimal set of stellar features. $P$ represents also the purity of the sample of accreted stars, while $R$ characterises the completeness of the sample. By definition, these two parameters correspond to the number of correctly and mis-predicted accreted stars, respectively:

\begin{equation*}
    P = \frac{TP}{TP+FP} \quad \textrm{and} \quad  R = \frac{TP}{TP+FN},
\end{equation*}

\noindent where $TP$, $FP$, and $FN$ are the number of true positives (i.e., stars that are correctly classified as accreted), false positives (in-situ stars which are misclassified as accreted), and false negatives (accreted stars which are misclassified as in-situ), respectively. 

Because the precision and recall values of both ANNs and decision tree-based models vary based on the different classification thresholds, with some thresholds favouring some models over the others, we also use as metric the area under the precision-recall curve, PR-AUC. This metric is more robust across models as it accounts for the precision and recall values evaluated on a range of thresholds common to all models. A random classifier would return a PR-AUC score equal to the fraction of accreted stars in the test dataset, whereas a perfect classification algorithm would have a PR-AUC score of 1.

We have elected not to employ the accuracy metric, defined as the overall fraction of correctly classified stars, due to its shortcomings for highly imbalanced cases.  For example, in galaxies with accreted fractions of $\approx$10\% (a typical value), a classifier which always predicts stars as being in-situ would have an accuracy of $\approx$90\% even though it failed to identify any accreted stars.
 
Aside from the common ML metrics, we also compare the models in terms of how well they fare in terms of astrophysical diagnostics. These are not metrics per se, however, they are useful to help understand whether the models are able to grasp the `physics' behind the data. We expect that a model that is able to learn (or mimic) the physical processes behind the origins of the two populations would be less precise exactly in those regions of the parameter space where the properties of the two populations are similar (for example, stars that were born in the early phases of the galaxy formation are old, metal-poor, more $\alpha$-enhanced, and move on less ordered orbits, regardless of whether they were born in-situ or accreted). A model that cannot learn the physical patterns may still have a good performance, however its mis-classifications may be distributed more randomly in physical space. The three diagnostics used here, are:  

\begin{itemize}
\item{[$\alpha$/Fe] -- [Fe/H] plane. Accreted stars tend to be located in the high [$\alpha$/Fe], low [Fe/H] region of the plane, while disc and in-situ halo stars generally have lower [$\alpha$/Fe] and higher [Fe/H] values.}
\item{Toomre diagram, which is the distribution of rotational velocity, $v_{\theta}$, versus the dispersion velocity $\sigma$. In this plane, the disc and the in-situ halo stars have high $v_{\theta}$ and low $\sigma$, whereas accreted stars do not have a preferred direction of motion, and generally have high $\sigma$.}
\item{$E-L_{z}$ distribution. As these parameters are quasi-conserved for a given infalling satellite, stars belonging to different disrupted progenitor would appear as ``clumps'' in this plane. These clumps are likely to be more distinct in the upper part of the plane, which is associated with late accretions. The in-situ halo stars, and the disc, are located on the region with positive $L_z$.}
\end{itemize}

\subsection{Supervised ML models}
\label{subsec:models}

\subsubsection{The benchmark model}
\label{subsec:benchmark}

For our analysis, we build a benchmark model to: i) investigate the most informative set of stellar properties that can distinguish accreted from in-situ stars, i.e., the \textit{optimal} set of features; ii) to compare the performance of different ML algorithms trained on this optimal set. 

Our benchmark model is similar to the Galactic Archaeology Neural Network (GANN) model of \citet{tronrud_machine_2022}. This is an MLP (see Section~\ref{subsec:MLP}) comprised of an input layer, a batch normalization layer, four hidden layers of 64, 256, 64, and 32 neurons, and an output layer with one neuron, resulting in a total of 35521 trainable parameters. We also use the same activation functions as in the GANN model in the corresponding layers. In the following, we refer to this configuration of layers and neurons as the benchmark architecture. 

The model is first trained on a set of stellar features comprising ages, [Fe/H] and [$\alpha$/Fe], as in the GANN model. Following the same approach as in GANN, the chemical abundances are expressed linearly (rather than the more conventional logarithmic form) and normalised by the correspondent solar values. However, unlike in GANN, we choose not to include the hydrogen fraction with [Fe/H], given that the information about the former can be implicitly reconstructed by the network from the latter. As for GANN, our benchmark model is trained on an equal number of accreted and in-situ stars. There are other differences from the GANN model also. One is that we train the model only on stars within $r\leq 50\, \textrm{kpc}$ from the centres of galaxies, as beyond this radius the distribution of stars is overwhelmingly of accreted origin (this is also valid for the other models). More significantly, we only include example stars from the main halo, whereas \citet{tronrud_machine_2022} include also those from present-day satellites, to augment the samples of accreted stars. Moreover, they adopt a strategy of drawing equal number of stars from satellites in different mass ranges, in order to increase the number of examples of stars from objects which contribute fewer stars (i.e., the low-mass dwarfs). In choosing to train only on existing debris, our benchmark model has a slightly lower performance than that of GANN, however the training data  represents a closer match to what one would expect from observations.

\begin{figure}
    \centering
    \includegraphics[width=\columnwidth]{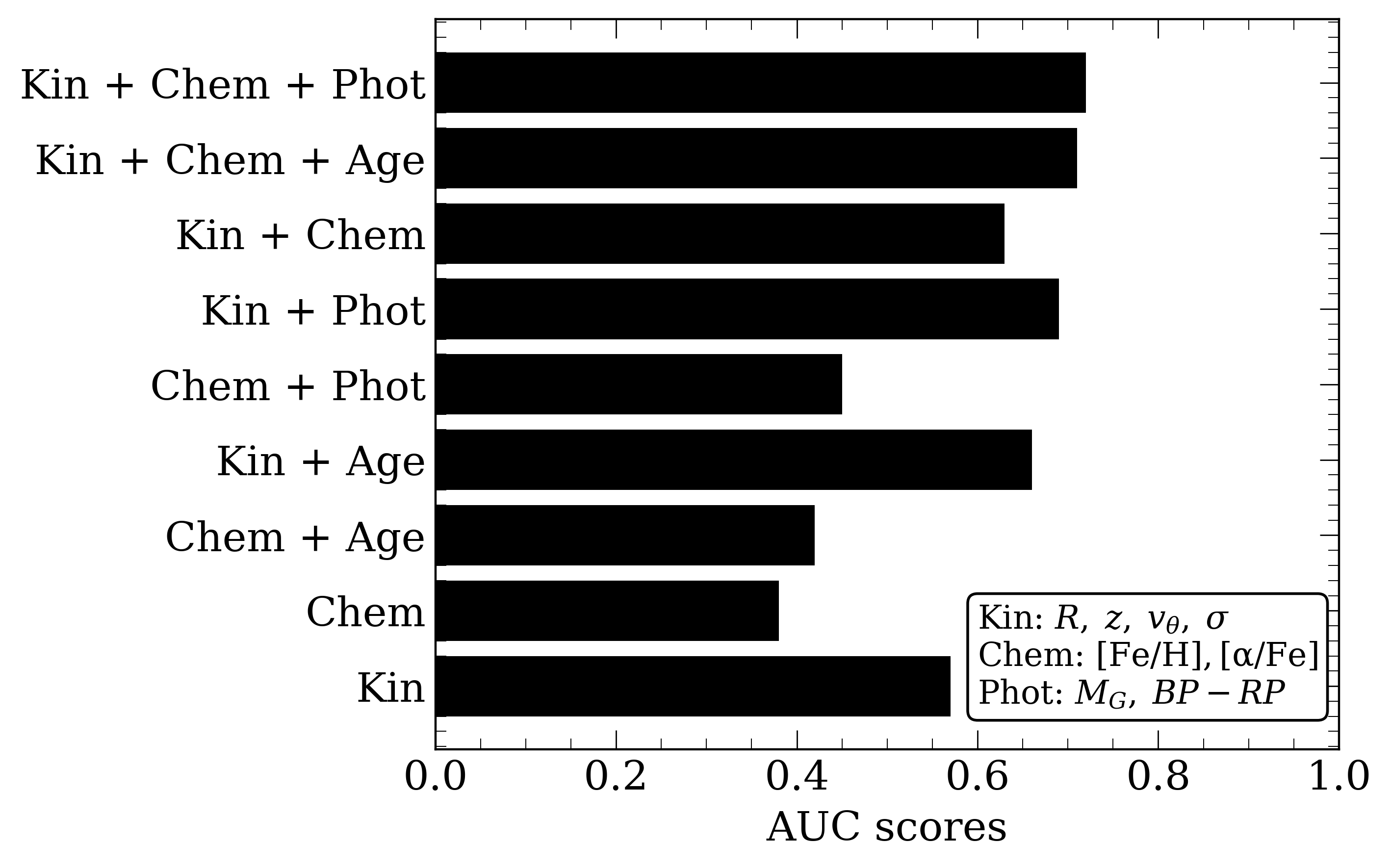}
    \caption{PR-AUC scores for the precision-recall curves for the models obtained training the benchmark architecture using different combinations of features. The feature categories are: positions and kinematics (labelled 'kin'), [Fe/H] and [$\alpha$/Fe] abundances ('chem'), ages, and {\it Gaia} magnitudes and colours ('phot').}
    \label{fig:features_selection}
\end{figure}

\subsubsection{The optimal set of stellar features}
\label{subsec:optimal_feat}

With the benchmark architecture, we proceed to compare the classification performance using different combinations of stellar features as inputs, to determine the optimal set for the models. We consider the stellar properties described in Section~\ref{sec:feature_selection}, which we divide into four categories: positions and kinematics ($R, \, z, \, v_{\theta}, \, \sigma$), chemical abundances ($\mathrm{[Fe/H]}, \, [\alpha/\mathrm{Fe}]$), ages ($\tau$) and photometry ($\mathrm{M}_{\mathrm{G}}, \, {\mathrm{BP}}-{\mathrm{RP}})$. For brevity, these are also referred to as 'kin', 'chem', 'age' and 'phot', respectively.

We then train the benchmark architecture with different combinations of these four categories and evaluate the classification performance on stars from the validation dataset. The results are shown in Fig.~\ref{fig:features_selection}.  On their own, positions and kinematics (kin) give a PR-AUC score of $0.57$, which is significantly higher than than the score for [Fe/H] and [$\alpha$/Fe] (chem), PR-AUC 0.38. This suggests that spatial distribution and kinematics are more informative than chemical abundances. In combination, kin + chem perform somewhat better than kin alone, with a PR-AUC score of 0.63, which suggests that the two categories provide complementary information. Adding ages to positions and kinematics improves the performance of the model compared to just adding chemistry (PR-AUC scores 0.66 for kin + age versus 0.63 for kin + chem). Similarly, adding ages to positions, kinematics and chemistry improves the score compared to not adding ages (PR-AUC of 0.71 for kin + chem + age compared to 0.63 for kin + age). Interestingly, photometry adds slightly more information than the ages, to both positions and kinematics (PR-AUC scores 0.69 for kin + phot versus 0.66 for kin + age), or chemistry categories (PR-AUC scores 0.45 for chem + phot versus 0.42 for chem + age). A possible explanation for this is that photometry is implicitly related to both the ages and metallicities of stars. 
 
Overall, the best performance is provided by kin + chem + phot (PR-AUC score of 0.72), therefore excluding ages. We therefore assign the corresponding input parameters, namely:
\begin{equation*}
\{R, \, z, \, v_{\theta}, \, \sigma, \mathrm{[Fe/H]}, \, [\alpha/\mathrm{Fe}],  \, \mathrm{M}_{\mathrm{G}}, \, {\mathrm{BP}}-{\mathrm{RP}} \}.    
\end{equation*}

\noindent as the \textit{optimal} set features for further training and comparison of the other  ML models.

We note, however, that the combinations kin + phot + age and kin + phot have classification performances which are very close to that of the optimal combination. This suggests that ML models could also be trained on reduced information, for example on just positions, kinematics and photometry, without a significant drop in performance. In fact, this may be a preferred option for observations, given that photometry is usually more readily available than spectroscopy.

\subsubsection{Multilayer perceptron}
\label{subsec:MLP}

A MLP is a type of ANN consisting of multiple layers of interconnected artificial neurons, or perceptrons. The architecture typically comprises an input layer, one or more hidden layers, and an output layer. Each neuron receives input signals from neurons in the previous layer and computes a weighted sum based on internal, tunable parameters describing the importance of the single inputs. Before being forwarded to the next layer, the result is passed to a non-linear activation function to allow the learning of non-linear relations between inputs and outputs. During the training process, the weights connecting the neurons are updated to minimise the error, as estimated by an objective function, between the predicted and actual class through an optimisation algorithm.

The benchmark model described earlier is also an MLP. However, because the classification between accreted and in-situ stars is inferred from a small number of features, we consider also shallower architectures than the one used in the benchmark model.
Specifically, we explore architectures comprised of one hidden layer of 10, 50, and 100 neurons, and two hidden layers with 50 neurons in each. In total, these four MLPs have 101, 501, 1001, and 3051 trainable parameters, respectively. In comparison, the benchmark architecture contains 35777 parameters.

We train the MLPs with various architectures on the optimal set of input features (described in Section \ref{subsec:optimal_feat}), as was done for the benchmark model. We adopt the same activation functions and optimisation algorithm for all cases. For the outputs of the neurons in the hidden layers, we apply a Scaled Exponential Linear Unit (SELU) function \citep{klambauer_self-normalizing_2017}, while for the neuron in the output layer we apply a sigmoid function, to ensure the prediction is in the range $0-1$. The trainable parameters are updated using the \texttt{Adam} \citep{kingma_adam_2017} optimisation algorithm on the error between predictions and labels estimated by the binary cross-entropy function. The training on the optimal set of features of the four MLP plus the benchmark architectures is done for a maximum of 100 epochs\footnote{An epoch is a complete pass of the MLP through all the examples in the training dataset.}, with an adaptive learning rate\footnote{The learning rate is the constant of proportionality relating the gradient of the objective function and the associated change in the trainable parameters of the MLP.} halving when the value of the objective function stops decreasing for more than 5 epochs. 

\begin{figure}
    \centering
    \includegraphics[width=0.8\columnwidth]{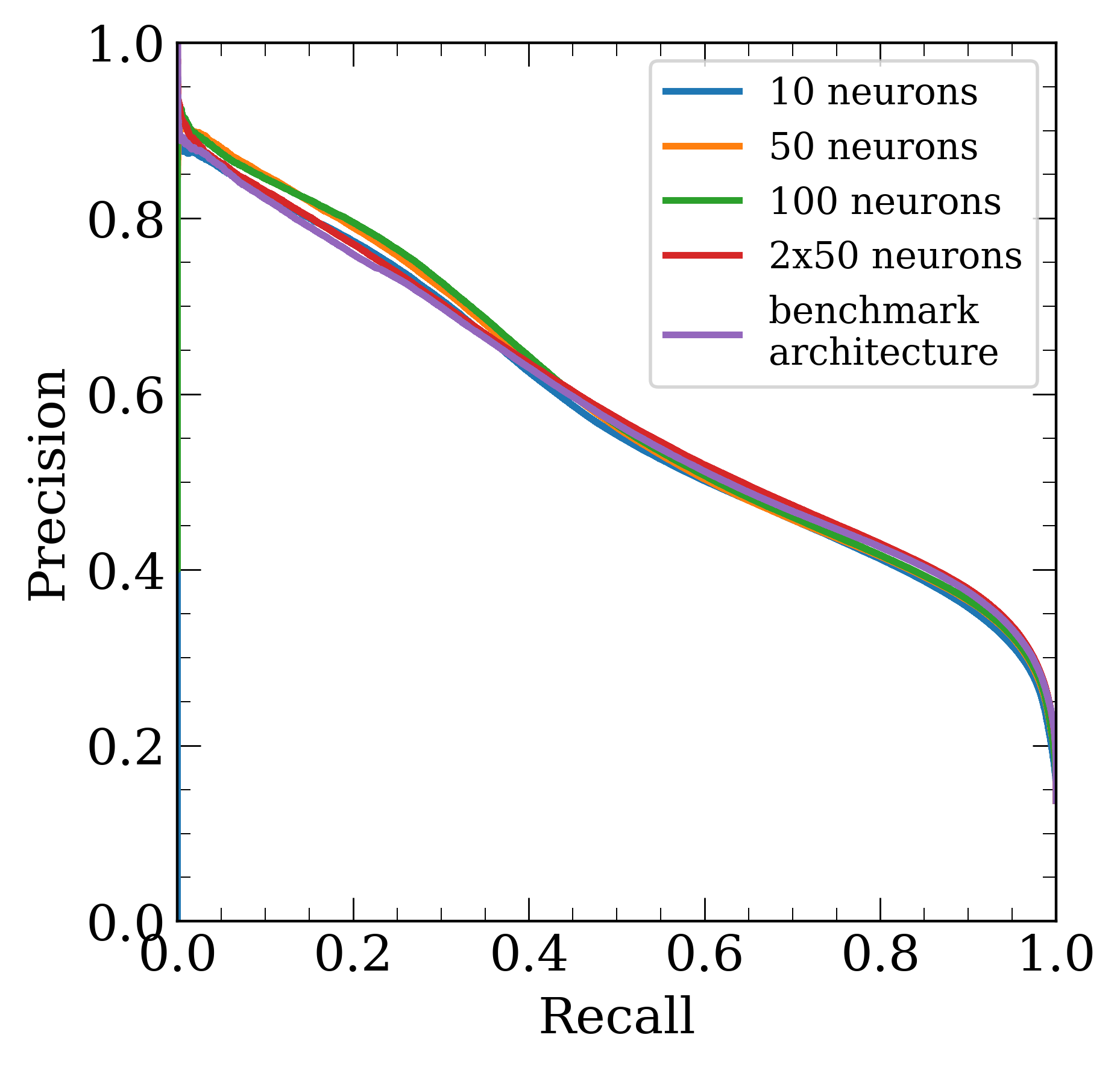}
    \caption{Comparison between the precision-recall curves obtained training the benchmark architecture and shallower ANNs on the optimal set of features (see Section \ref{subsec:optimal_feat}).}
    \label{fig:mlp_acrhitecture}
\end{figure}

Fig.~\ref{fig:mlp_acrhitecture} shows a comparison of the classification performances of all these architectures, based on the PR-AUC score evaluated on the test dataset. All MLPs have similar performances, with PR-AUC scores ranging from 0.578 (1 hidden layer, 10 neurons model) to 0.591 (1 hidden layer, 100 neurons model). Despite the significantly larger number of trainable parameters, the benchmark architecture (0.584) outperforms only the shallowest model. The model with 1 hidden layer and 50 neurons, and the one with 2 hidden layers of 50 neurons each, have PR-AUC score of 0.589 and 0.590, respectively. As the architecture comprising 1 hidden layer with 100 neurons returns the highest PR-AUC score (0.591), we considered it for the rest of the analysis (hereafter denoted as MLP). 

\begin{table*}
    \centering
    \begin{tabular}[H]{ccccccccccccc}
        \hline
         & G01 & G15 & G17 & G18 & G19 & G23 & G24 & G25 & G27 & G38 & G40 & G44\\
         \hline
         G01 & 0.68 &  0.60	& 0.52&	0.66&	0.44&	0.60&	0.59&	0.45&	0.81&	0.48&	0.76&	0.66\\
         
         G15&	0.54&	0.77&	0.48&	0.67&	0.58&	0.62&	0.67&	0.44&	0.74&	0.62&	0.66&	0.66\\

         G17&	0.58&	0.64&	0.62&	0.62&	0.62&	0.70&	0.61&	0.47&	0.73&	0.59&	0.67&	0.71\\

         G18&	0.55&	0.66&	0.45&	0.76&	0.46&	0.50&	0.65&	0.46&	0.78&	0.48&	0.74&	0.57\\
         
        G19&	0.54&	0.67&	0.47&	0.64&	0.74&	0.71&	0.66&	0.44&	0.73&	0.62&	0.66&	0.65 \\
        
        G23&	0.57&	0.67&	0.52&	0.62&	0.69&	0.82&	0.61&	0.44&	0.76&	0.62&	0.63&	0.72\\
        
        G24&	0.52&	0.66&	0.43&	0.67&	0.48&	0.56&	0.78&	0.45&	0.79&	0.53&	0.75&	0.61 \\
        
        G25&	0.47&	0.43&	0.44&	0.58&	0.44&	0.46&	0.39&	0.61&	0.62&	0.32&	0.46&	0.58 \\
        
        G27&	0.48&	0.49&	0.35&	0.63&	0.25&	0.36&	0.47&	0.42&	0.88&	0.53&	0.76&	0.50 \\
        
        G38&	0.51&	0.62&	0.47&	0.63&	0.45&	0.50&	0.52&	0.45&	0.77&	0.76&	0.70&	0.65\\
        
        G40&	0.47&	0.43&	0.33&	0.63&	0.16&	0.29&	0.43&	0.39&	0.78&	0.29&	0.83&	0.46\\
        
        G44&	0.57&	0.64&	0.55&	0.63&	0.64&	0.65&	0.63&	0.46&	0.78&	0.58&	0.73&	0.83\\


        \hline

    \end{tabular}
    \caption{PR-AUC scores for the MLP models trained on accreted and in-situ examples from the galaxy in the left-most column, and tested on the galaxies listed in the top row. Where the galaxy label is the same, stars in the validation dataset were considered.}
    \label{tab:mlp_single_halos}
\end{table*}

\subsubsection{Multilayer perceptron with galaxy features}
\label{sec:MLP_galaxyfeatures}

The study of \citet{tronrud_machine_2022} suggests that MLP models may be biased towards specific assembly histories. This result is expected, especially when the training set does not contain sufficient types of accretion histories. We find a similar result when we train a MLP model on a single galaxy. As expected, the model performs better when the stars belong to the same galaxy. 
Table \ref{tab:mlp_single_halos} shows the classification performances of these MLP models represented by the PR-AUC scores. The models are trained on the optimal set of features using accreted and in-situ examples from one galaxy (listed in the left-most column) and tested on another galaxy (listed in the top row). Where the galaxy pairs are the same, we only use the stars in the validation dataset.  On a galaxy basis (i.e., column-by-column analysis in the table), the best classification performance is always associated to the model explicitly trained on the galaxy it is tested on. 

This result is a consequence of the unique assembly history of each galaxy, where the properties of the progenitors, e.g., their infall times, orbits, and masses, imprint a specific characteristic in the stellar properties (features) of accreted stars. For instance, in galaxies where the MMAP was accreted at early times the accreted stars are mostly rich in $\alpha$-elements and are more phase-mixed; conversely, in galaxies with a late MMAP, the accreted stars tend to be more $\alpha$-poor and more spatially coherent \citep{Font2006a}. 
Consequently, the MLP model learns the details of the unique imprint of the assembly history of a given galaxy leaves in its distribution of accreted stars. 

We note, however, that some models are able to identify a purer and more complete sample of accreted stars in other galaxies than the one they had been trained on. Again, this can be directly related to a difference in assembly histories as some galaxies (e.g., G27) have a neater distinction between the accreted and in-situ populations.

To account for the variability in the properties of accreted stars in different systems, we could choose to use a conglomeration of galaxies in the training set, as in the dataset described in Section \ref{sec:data}.  However, this model would still underperform if applied onto a galaxy whose assembly history is not represented in the training set. In ML terminology, the performance degradation of a model trained on a source distribution (``domain'') when applied to a statistically different target one is referred to as ``domain shift'' \citep{amos_when_2008}. 
Since our aim is to create a model capable of generalising across different assembly histories, we retrain the MLP model by providing additional information about the galaxy from which the example stars are taken from. These additional input features are global properties of host MW-mass galaxies, measured within an aperture of $30$~kpc. This approach has the advantage of providing information about the accretion history of the host galaxy, without given the model any a priori knowledge of which stars were accreted. The galaxy-specific input features are described in Section~\ref{sec:galaxy_feature_selection} and are listed in Table~\ref{tab:datasets}.

\subsubsection{Transformational machine learning}
\label{subsec:tml}

As an alternative method to mitigate the domain shift problem, we consider the TML technique of \citet{olier_transformational_2021}. In this framework, each data point is described by a vector of predictions obtained from an ensemble of \textit{base} models. While using a common set of features, the base models are sequentially trained on different examples to perform different tasks. This representation of the data is used as an input to a new model, which combines the prior information encoded in the base models. 

Before the implementation of the TML approach, we trained an MLP model (see Section~\ref{subsec:MLP}) on each galaxy in the training dataset. Because of the differences in the assembly histories of these galaxies, learning to classify the accreted stars is considered by the model to be a specific task for each galaxy.  The resulting ensemble of MLPs is then used to derive a predictive description of all the stars in the training dataset. These 12D vectors are then passed as inputs to a single-layer ANN with 100 neurons. During training, the internal parameters of the base MLP models are held constant, while the parameters from the neural network which combines the predictions are allowed to update. Also, to avoid any data leakage, the predictions of a base model on the stars used for the training are set to $0$.

\subsubsection{XGBoost}
\label{subsec:xgb}

A decision tree is a ML algorithm based on a tree-like structure. It begins with a root node representing the entire dataset and recursively splits the data into smaller subsets (branches) based on feature values. The endpoint of a branch is called a \textit{leaf} and contains the model prediction. In this work, we combine many decision tree models trained on the same dataset using the gradient boosting machine (GBM) method \citep{friedman_greedy_2001}. Following this algorithm, a decision tree is created to separate accreted and in-situ stars by predicting their classification label as a continuous score between 0 and 1. Then, a new decision tree is added to predict the error between the predicted and actual labels (here, as for the MLP model, estimated by the binary-loss function). The prediction from the new model is then added to the initial predictions to make a more accurate classification. This continues for an arbitrary number of iterations, with each new model sequentially added and trained to minimise the error of the whole ensemble.

Here we implement a GBM with decision tree models using the XGBoost \citep{chen_xgboost_2016} method. This algorithm is particularly suitable for large datasets as the ensemble of models is built in parallel rather than serially. Moreover, it includes L1 and L2 regularization techniques to control over-fitting. The number of base models in the ensemble and the number of splits in each tree are decided using the \textsc{optuna}\footnote{https://optuna.org/} hyperparameter optimisation framework.

\subsubsection{UMAP}
\label{subsec:umap}

The UMAP method \citep{mcinnes_umap_2018} is a dimensionality reduction  technique, such as Principal Component Analysis, with the advantage that the obtained dimensions can be non-linearly related to the starting ones. We apply this method in order to find relations, or identify possible structures within the accreted and in-situ examples that may exist in the training and test datasets. Given the specificity of this method, we use it only for visualising the data, and therefore we do not include it in our classification performance comparison.  

Assuming the data are uniformly distributed on a locally connected Riemannian manifold, the algorithm constructs a fuzzy topological structure of it in a $8$D parameter space and maps it into a lower dimensional space with the closest equivalent structure. UMAP can be used in a supervised way by providing the labels of the classes into which the data are separated. This ensures that both the global and class-specific structures of the data are retained while maximising their separation in the new embedding.  

Here we train a UMAP model to reduce the 8D parameter space defined by the optimal set of stellar features into a 2D plane, where the separation between the accreted and in-situ training examples is maximised. Because of the high-computational resources required by this method, we consider only a subset of examples from each galaxy, which consists of all the accreted examples and an equal amount of the in-situ ones. This results in a  statistically significant number of examples from each galaxy. To investigate potential differences between the structures of the training and test datasets, we use the same UMAP model to project all the test examples into the 2D plane. A discussion of the results of this UMAP embedding is provided in Section \ref{subsec:umap_results}.

\section{Model comparison}
\label{sec:results}

\subsection{Classification performance}
\label{subsec:comparison_ML}

With the optimal set input features, we proceed to compare the classification performance of our models, MLP, MLP with galaxy features, TML and XGBoost, in separating the two classes of stars. We evaluate all models on the test dataset, and compare the classification performances using the PR-AUC scores. We also assess the purity and completeness of the retrieved accreted samples using the precision and recall metrics at the optimal threshold value. To ensure a fair comparison, for each model we consider the threshold associated to the highest value of the harmonic mean between precision and recall (i.e., the F1-score). The resulting fiducial thresholds are shown in Table \ref{tab:fiducial thresholds}.

\begin{figure}
    \centering
    \includegraphics[width=0.85\columnwidth]{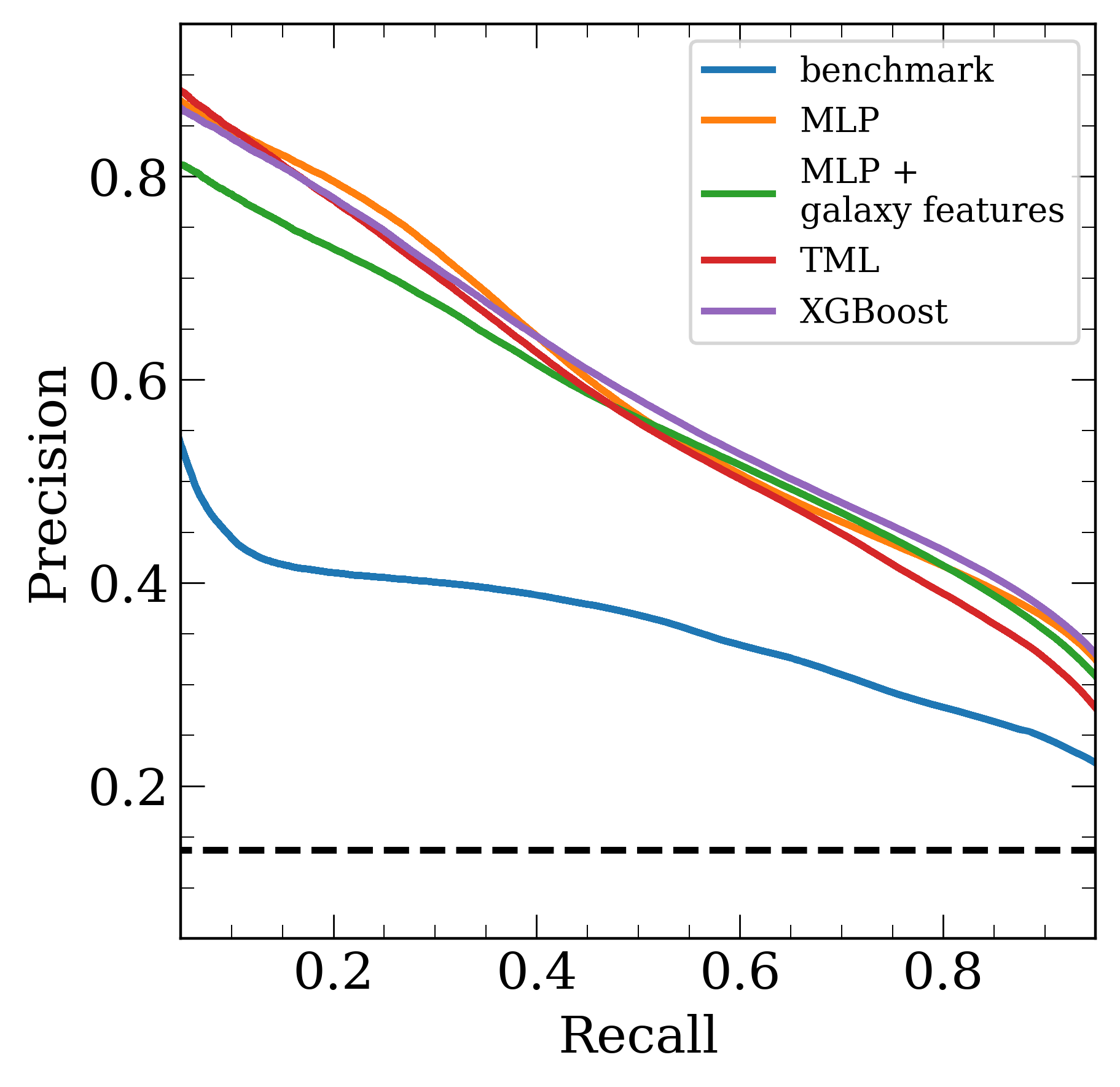}
    \caption{Precision and recall values at different classification thresholds for the models. The metrics were evaluated considering all the stars in the test dataset.}
    \label{fig:PR_test}
\end{figure}

\begin{table}
    \centering
    \begin{tabular}{l|c|c}
    \hline
        Model &  Threshold & F1-score \\
        \hline
         MLP & 0.33 & 0.67\\
         
         MLP + & \multirow{2}*{0.31} & \multirow{2}*{0.63}\\
         galaxy features & & \\
         
         TML & 0.24 & 0.62\\

         XGBoost & 0.33 & 0.68\\
         \hline
    \end{tabular}
    \caption{Fiducial classification threshold values for the models. Each value is associated to the highest F1-score calculated based on the precision and recall values on the validation dataset.}
    \label{tab:fiducial thresholds}
\end{table}

\begin{table*}
    \centering
    \begin{tabular}{c|c|c|c|c|c|c|c|c}
      \hline
      Model   & \multicolumn{2}{|c|}{G29} & \multicolumn{2}{|c|}{G30} & \multicolumn{2}{|c|}{G34} & \multicolumn{2}{|c|}{G42}\\
       & $P$ & $R$ & $P$ & $R$ & $P$ & $R$ & $P$ & $R$ \\
      \hline
      benchmark & 0.15 & 0.98 & 0.45 & 0.91 & 0.09 & 1.00 & 0.21 & 0.97 \\

      MLP & 0.47 & 0.64 & 0.59 & 0.57 & 0.35 & 0.65 & 0.54 & 0.54 \\

      MLP + & \multirow{2}*{0.54} & \multirow{2}*{0.59} & \multirow{2}*{0.56} & \multirow{2}*{0.56} & \multirow{2}*{0.41} & \multirow{2}*{0.55} & \multirow{2}*{0.50} & \multirow{2}*{0.63} \\
      galaxy features & & & & & & & & \\

      TML & 0.46 & 0.73 & 0.58 & 0.55 & 0.32 & 0.72 & 0.54 & 0.60 \\

      XGBoost & 0.49 & 0.70 & 0.58 & 0.58 & 0.36 & 0.68 & 0.54 & 0.58 \\

      \hline
    \end{tabular}
    \caption{Comparison of the precision and recall values evaluated on the test dataset at the fiducial thresholds for the ML models.}
    \label{tab:ml_models comparison}
\end{table*}

Fig.~\ref{fig:PR_test} shows the comparison between these four models, plus the benchmark model. For the latter, we use the reduced set of features described in Section~\ref{subsec:benchmark}. This figures shows that all four models perform significantly better than the benchmark, which indicates the importance of adding more input parameters, in this case from the kinematical and photometric properties of the stars. The MLP, TML and XGBoost models perform similarly on the test dataset, with PR-AUC scores of 0.59, 0.57, and 0.59 respectively. The TML model does  not provide any improvement over the MLP, which implies that there is not enough variance among the single MLPs in the ensemble.  

The MLP model with added galaxy-specific features gives a PR-AUC score of 0.57, which is worse than the one of the MLP model and the one calculated on validation data (0.64). This suggests that the galaxy-specific features favour the learning of specific patterns in the data, leading to overfitting, rather than learning the distinction between the accreted and in-situ classes in those features. This could be due to redundancies in the set of galaxy-specific features (which, unlike the stellar features, have not been optimised). For example, information gained from galaxy stellar masses may be very similar to that inferred from $\langle$[Fe/H]$\rangle$, as MW-mass systems follow the stellar mass -- metallicity scaling relation. Also, the information gained from the overall rotation of stars (i.e., $\kappa_{\rm co}$) may overlap with that obtained from the sizes of galaxies. In the future, we aim to investigate whether an optimal set of galaxy-specific features exists, particularly one that will increase the performance above the model without any such features. Alternatively, it could be that the galaxy properties used for this task are not representative of the specific accretion histories. In this case, one may opt, instead, to use parameters more directly related to the merger histories, for example the properties of the MMAPs. This will be investigated in a future study.

Table~\ref{tab:ml_models comparison} shows a galaxy-by-galaxy comparison of all five models based on the $P$ and $R$ values evaluated at the fiducial classification thresholds. Confirming what was found previously, MLP, TML and XGBoost have similar performances for every test galaxy,  consistently retrieving $>50\%$ of accreted stars with similar purity. In general, the TML model retrieves a larger sample of accreted stars than the single MLP, with only a minor decrease in precision.

Overall, despite belonging to different families of ML methods, the MLP, TML and XGBoost models show similar performances, suggesting that the same underlying relations between stellar properties and their origin are learned from the data. This is an encouraging result for further applications. 

\subsection{Model comparison on physical diagnostics}
\label{subsec:fp_fn}

So far, we have shown that all ML models trained on the optimal features perform similarly well, which is remarkable given that accreted stars comprise only a small fraction of the total stellar budget. However, despite the similar overall performance, the models may still perform differently in certain regions of the physical parameter space covered by the accreted/in-situ stars. If the models are truly able to extract the physical properties of the two stellar populations, we expect that they will perform better in regions where the two populations are clearly distinct in physical parameters (e.g., kinematics, or metallicity) and less well in regions where these properties overlap. To investigate this possibility, we map the distribution in a chemo-dynamical parameter space of false positives (in-situ stars misclassified as accreted) and of false negatives (accreted stars misclassified as in-situ) in different models. In Section~\ref{subsec:metrics} we introduced several physically-motivated chemo-dynamical diagnostics, such as the Toomre diagram, or the [$\alpha$/Fe] -- [Fe/H] and $E-L_z$ planes, which we use here to evaluate how well can the models identify accreted stars in these parameter spaces.

\begin{figure*}
    \centering
    \includegraphics[width=1.0\textwidth]{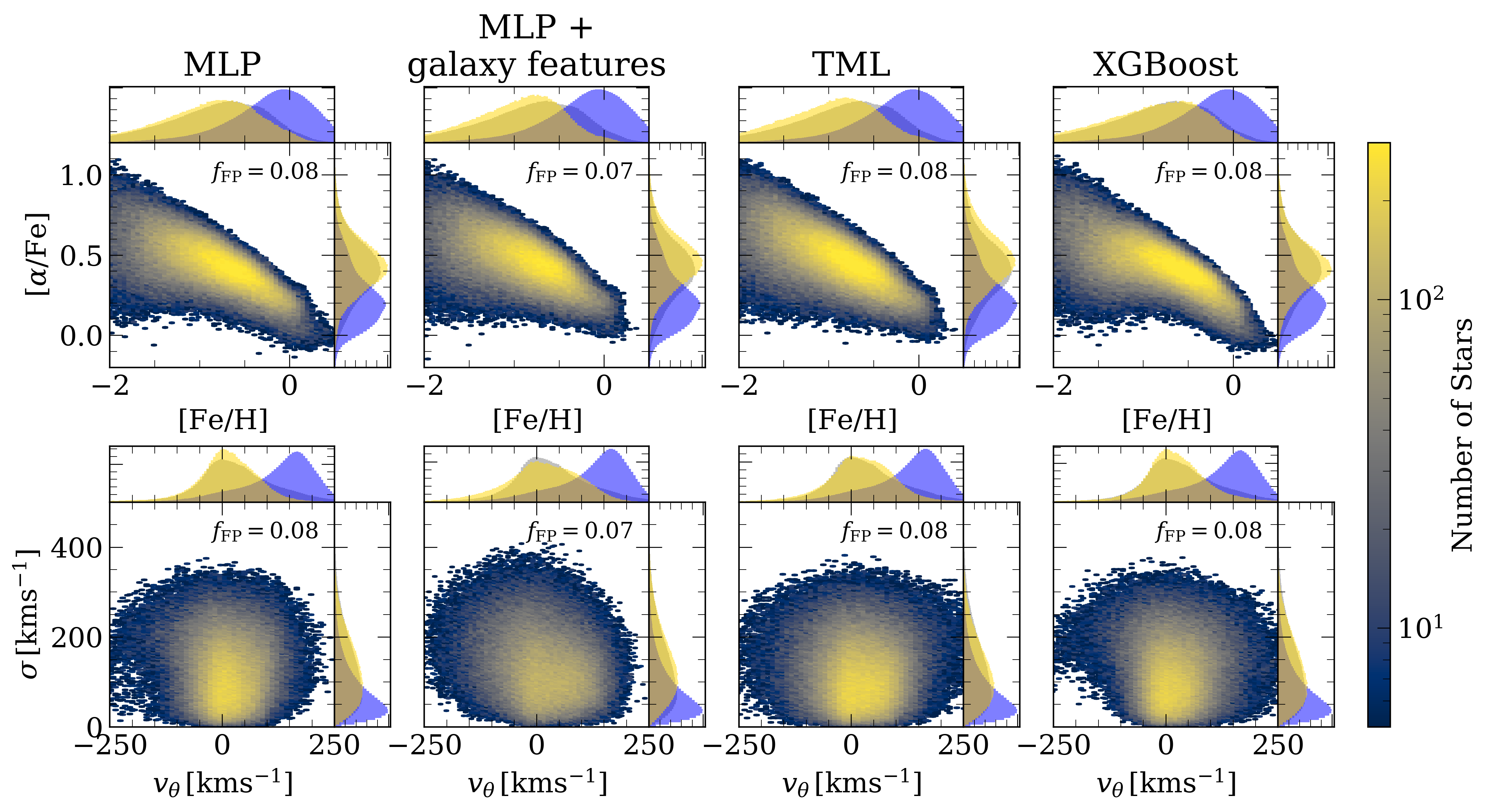}
    \caption{Distribution of the FPs in the test set in the [$\alpha$/Fe] -- [Fe/H] plane (top row) and in the Toomre diagram (bottom row). Columns from left to right correspond to the MLP, MLP+galaxy features, TML and XGBoost models, respectively. For each panel, the top and side sub-panels show the probability density function of the FP distributions (yellow) and of the accreted (grey) and in-situ (blue) training examples. For each model, we also show the FP fraction ($f_{FP}$) of the total number of stars in the test dataset. The metrics are evaluated at the fiducial threshold values listed in Table~\ref{tab:fiducial thresholds}.}
    \label{fig:FP}
\end{figure*}

\begin{figure*}
    \centering
    \includegraphics[width=1\textwidth]{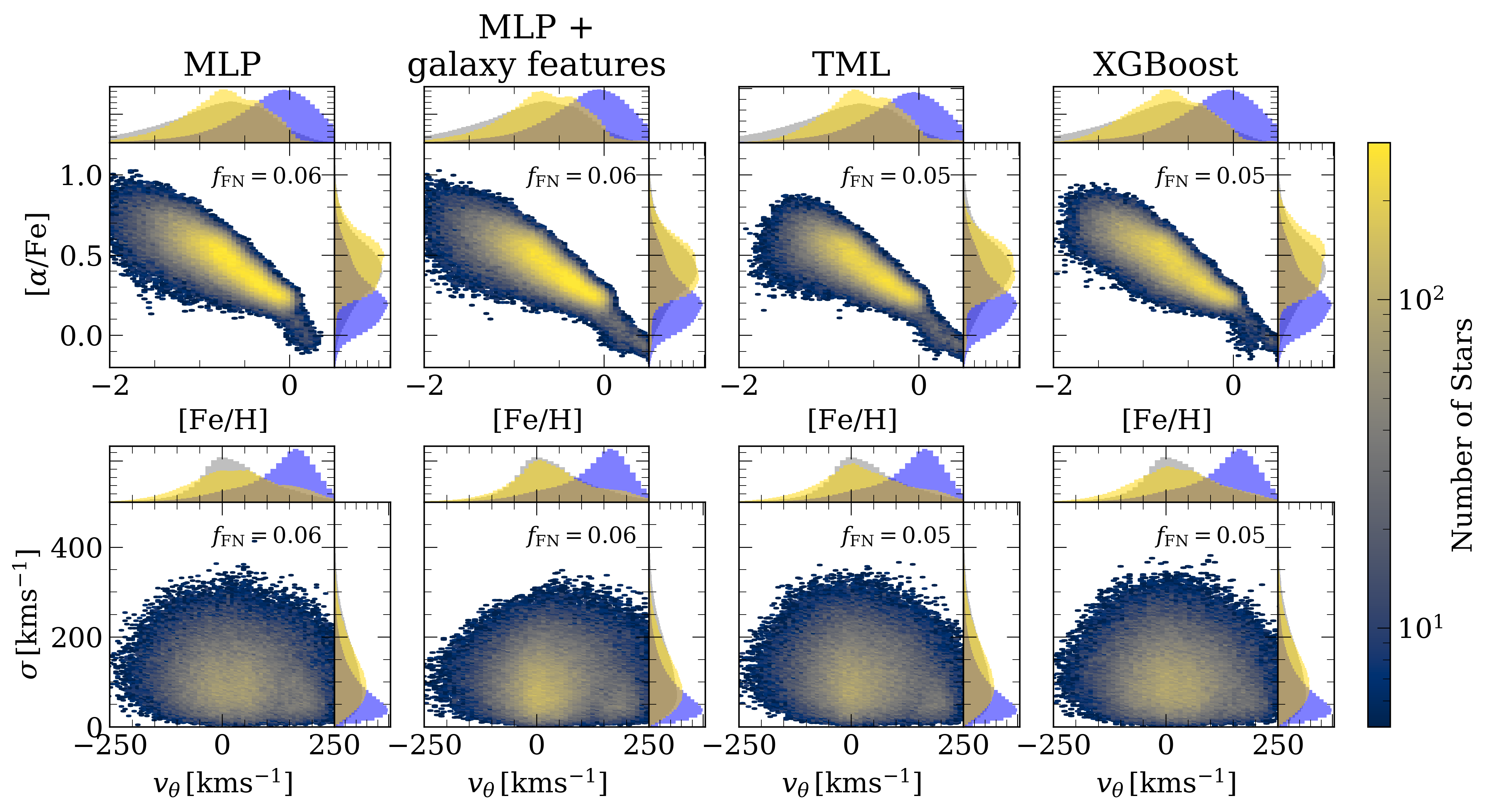}
    \caption{Same as in Fig.~\ref{fig:FP}, but for the accreted stars mis-classified as in-situ, i.e., the false negatives.}
    \label{fig:FN}
\end{figure*}

Fig.~\ref{fig:FP} shows the distribution of the FPs in the entire test galaxy set in the [$\alpha$/Fe] -- [Fe/H] plane (top row) and in the Toomre diagram (bottom row), for four models: MLP, MLP+galaxy features, TML and XGBoost, respectively. Each panel has a corresponding set of top and side sub-panels, in which we compare the probability density functions of the FPs (shown in yellow) with that of the accreted stars (grey), and of the in-situ stars (blue) in the training set. This figure shows that the FPs tend to be located in regions of the parameter space with low rotational velocities ($v_{\theta} \simeq 0$) and lower metallicities ([Fe/H] $<-1$), which are regions dominated by accreted stars. This indicates that the models can learn to identify accreted stars as typically more metal-poor and characterised by more chaotic motion, which are properties expected for this category. However, models find it more difficult to identify in-situ stars in this chemo-dynamical region. This behaviour is seen across all four models, which indicates that they all learn similar physical patterns for accreted stars. This result suggests that the similar classification performances of the models, found earlier, are in fact physically motivated. 

 
Similarly, in Fig.~\ref{fig:FN} we investigate the distribution of FNs in the same chemo-dynamical parameter space. 
This figure shows that, although the MLP, TML and XGBoost models retrieve the majority of the accreted stars in the test galaxies (see Table \ref{tab:ml_models comparison}), some accreted stars are still missed, despite them having relatively distinct motions and chemical abundance distributions from those of the in-situ stars (compare, again, the yellow, grey and blue probability distribution functions). All models present a similar behaviour in this respect, as was the case for FPs. We note here, too, that the FNs represent only a small fraction of the total number of stars, with $f_{\rm FN} \approx 5-7\%$ across different models.

The mis-classification in the case of FNs is likely due to the in-situ stars greatly outnumbering the accreted stars in the examples available to these models. To elucidate this, we also investigate the spatial location of the FNs. For the MLP, TML and XGBoost models, we find that the majority of mis-classified accreted stars (76\%, 77\% and 73\%, respectively) lie within a galactocentric radius of $5\,\textrm{kpc}$. This suggests that the models tend to identify more accurately the stars originating from late accretions, which are generally located in the outer regions of galaxies, while struggling to retrieve the stars that originate from early accretion events and which are now fully phase-mixed in the inner region. A possible solution to improving the classification of models can be provided by data augmentation techniques, which can be used to generate a higher number of accreted stars in the training sets in the inner region. Outside the $5\, \textrm{kpc}$ range, the MLP, TML and XGBoost models identify $93\%$, $94\%$, and $91\%$ of the accreted stars in the test dataset, with a precision of $0.57$, $0.56$, and $0.59$ at the fiducial classification threshold, respectively.

\begin{figure}
    \centering
    \includegraphics[width=0.9\columnwidth]{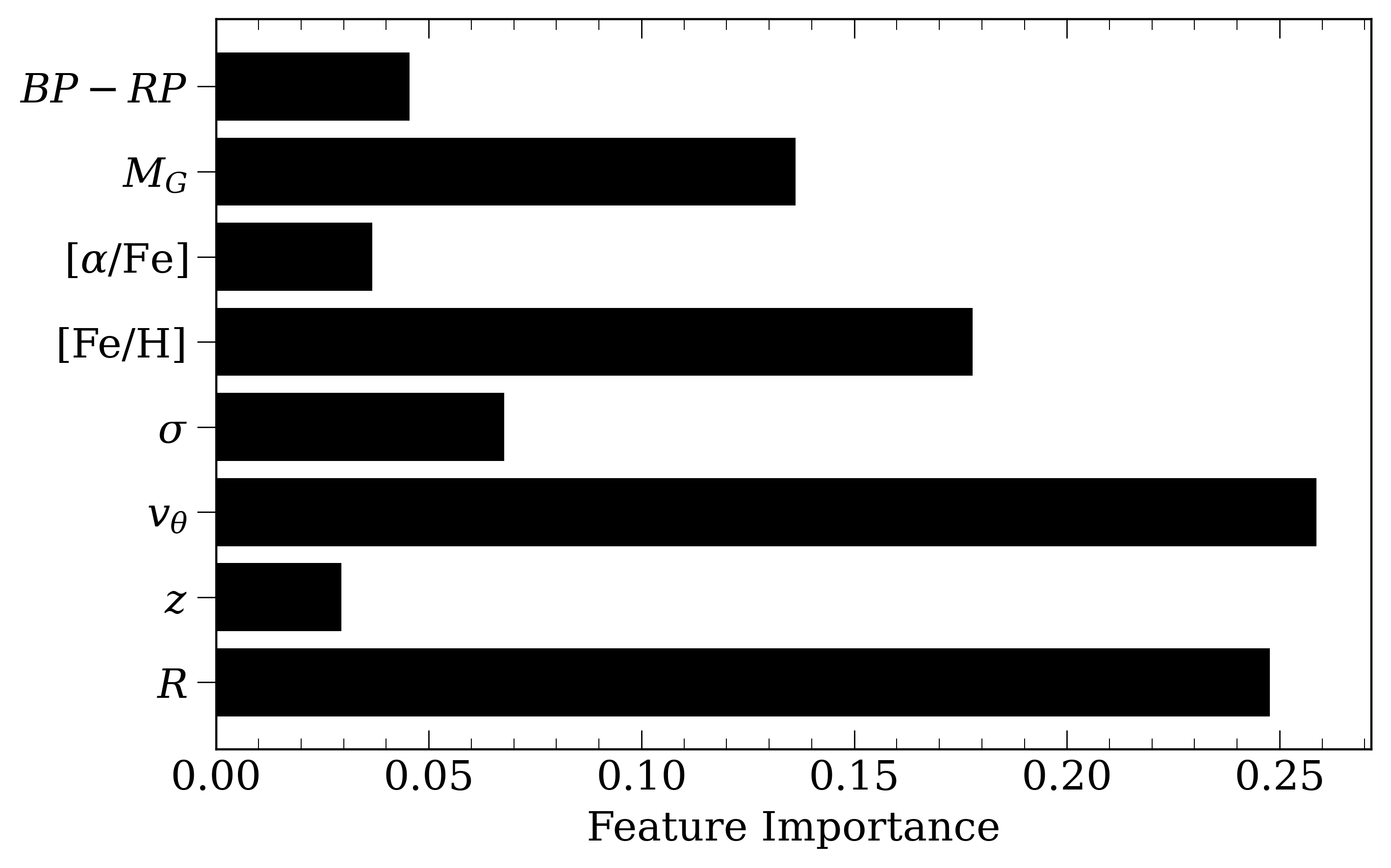}
    \caption{Information gain values describing the importance of the input features used by the XGBoost model to distinguish between accreted and in-situ stars. }
    \label{fig:xgb_features_importance}
\end{figure}

Furthermore, since the XGBoost model has a built-in degree of explainability, it can be used to determine more quantitatively the contribution of each input feature during the classification. Specifically, at each decision node, it is possible to calculate the \textit{information gain} from a given feature by subtracting the impurity (i.e., a measure of the entropy in the tree) before and after the splitting. The information gain of a specific feature in the XGBoost model can be estimated by averaging over all trees in the ensemble. Fig.~\ref{fig:xgb_features_importance} shows that the rotational velocity ($v_{\theta}$) and the distance from the centre of the galaxy ($R$) are the most important parameters for distinguishing between the accreted and in-situ stars in this model. Surprisingly, [$\alpha$/Fe] has a significantly lower information gain. Given the high importance of the [Fe/H] parameter, it is possible that the model considers the information provided by [$\alpha$/Fe] as redundant. Therefore, although the abundance of $\alpha$-elements can be used to characterise individual accreted substructures, the information from [Fe/H] may be sufficient to remove the in-situ background. 

\begin{figure*}
    \centering
    \includegraphics[width=0.7\textwidth]{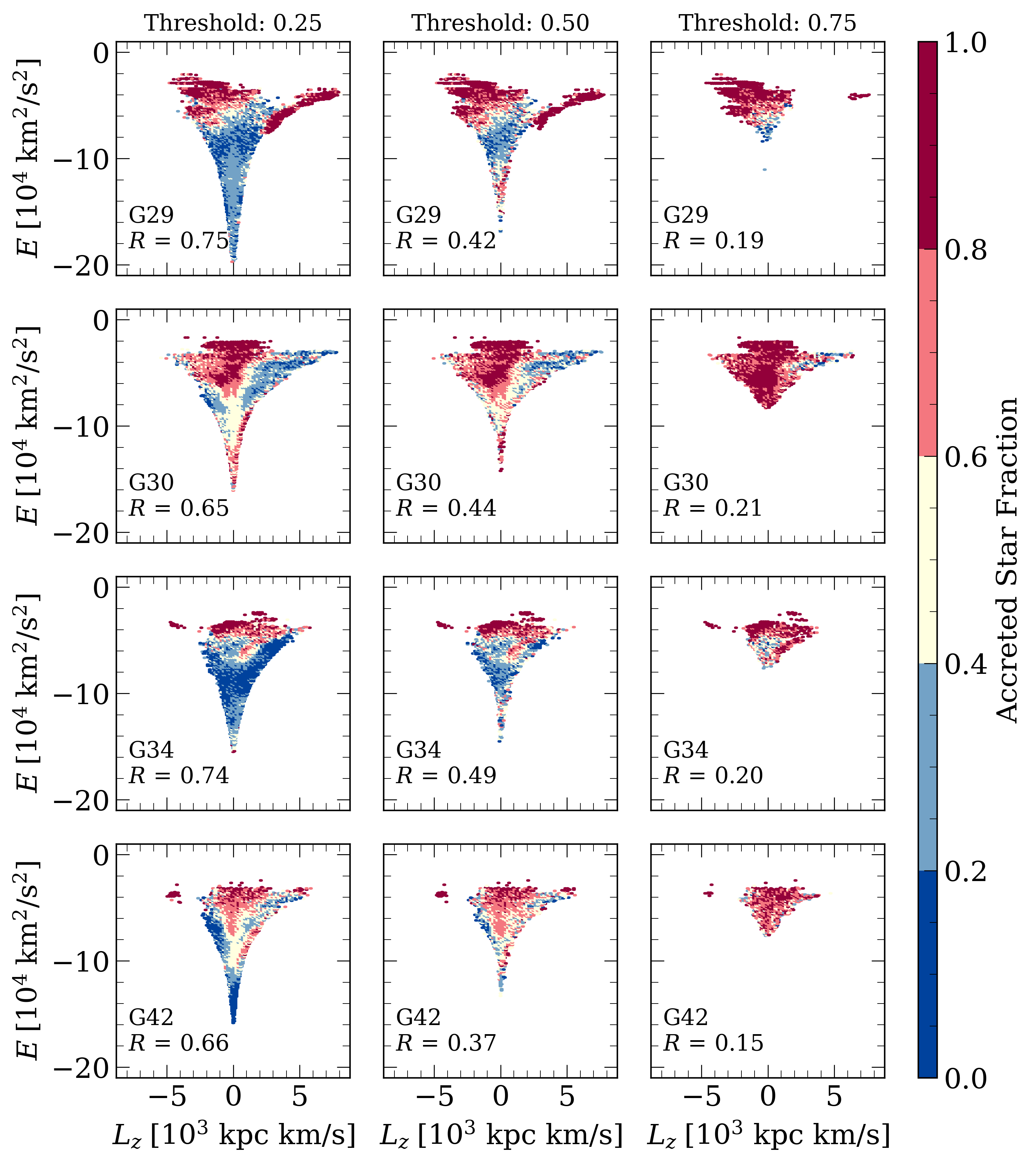}
    \caption{The distribution in the $E-L_{z}$ plane of the accreted sample retrieved by the MLP model, at different classification thresholds (left to right panels),  for each of the galaxy in the test dataset (top to bottom rows corresponding to galaxies G29, G30, G34, and G42, respectively.). The completeness ($R$) of the retrieved sample is also reported. The distribution is colour-coded by the actual fraction of accreted stars as defined by the simulation label.}
    \label{fig:diff_thresh_energy}
\end{figure*}

We note that the model performances reported here depend on the chosen classification threshold. For example, for the MLP model, the average purity of the accreted stars sample at the fiducial threshold is  $\simeq50\%$ (see Table~\ref{tab:ml_models comparison}). Purer samples can be obtained by increasing the classification threshold, however, this is done at the cost of completeness. This is shown in Fig.~\ref{fig:diff_thresh_energy}, where we plot the predictions of this model in the $E-L_z$ plane, for different threshold values (0.25, 0.5 and 0.75, respectively). Here $E$ denotes the total energy of a star, composed of the sum of kinetic and potential energy, while colours indicate the stellar mass fraction of accreted stars, as predicted by the model. The results are shown individually, for the four galaxies in the test dataset. 
The regions of the accreted components which are identified with more confidence by the models are those in the upper parts of the energy spectrum, which represent mostly stars originating from late accretions. These regions are dominated by clumpy structures which correspond to tidal debris not yet fully mixed with the rest of the halo. 

This result is encouraging and indicates that ML models are sensitive to physical patterns in the data, and may be used in the future to not only identify the bulk of accreted stars in the halo, but also to find individual tidal streams. Adjusting the classification threshold could provide an advantage compared with traditional methods of selecting halo stars which are fixed (see next section), whereas ML models can be customised to provide the most appropriate samples for different types of analysis. For instance, the identification of accreted substructures in integrals-of-motion space requires a sample of high purity to avoid the identification of spurious clusters, whereas a characterisation of substructures based on a large number of chemical abundances may be conducted at a lower purity, as the contamination of in-situ stars can be more easily identified. In this case, the model would represent a pre-processing step for reducing the number of stars to analyse.

\subsection{The ML performance in separating components versus observational cuts}
\label{subsec:selection_cuts}

\begin{table*}

    \centering
    \begin{tabular}{l|c|c|c|c|c|c|c|c|c}
      \hline
      & Halo (Accreted)   & \multicolumn{2}{|c|}{G29} & \multicolumn{2}{|c|}{G30} & \multicolumn{2}{|c|}{G34} & \multicolumn{2}{|c|}{G42}\\
      & selection criterion & $P$ & $R$ & $P$ & $R$ & $P$ & $R$ & $P$ & $R$ \\
      \hline
         
      \citet{helmi_merger_2018} & $\left| \boldsymbol{V} - \boldsymbol{V_{LSR}} \right| > \mathrm{210} \, \mathrm{km}\mathrm{s}^{-1}$ & 0.28 & 0.81 & 0.47 & 0.83 & 0.21 & 0.79 & 0.47 & 0.54 \\

      \citet{myeong_halo_2018} & $[\mathrm{Fe}/\mathrm{H}]<-0.5 \, \land \, v_{\theta} < 150 \, \mathrm{km}\mathrm{s}^{-1}$ & 0.28 & 0.93 & 0.49 & 0.65 & 0.09 & 0.87 & 0.48 & 0.76 \\

      \citet{massari_origin_2019} & $\varepsilon>0.5$ \ & 0.28 & 0.65 & 0.48 & 0.67 & 0.21 & 0.68 & 0.49 & 0.52 \\
      \hline
      MLP (0.33, fiducial) & & 0.35 & 0.75 & 0.52 & 0.71 & 0.22 & 0.93 & 0.51 & 0.56 \\
      MLP (0.10) & & 0.27 & 0.98 & 0.46 & 0.97 & 0.11 & 1.00 & 0.48 & 0.95 \\
      MLP (0.75) & & 0.70 & 0.03 & 0.88 & 0.05 & 0.55 & 0.04 & 0.76 & 0.02 \\
      \hline
    \end{tabular}
    \caption{Comparison of the purity ($P$) and completeness ($R$) of the sample of accreted stars retrieved by using observational selection cuts (top three rows) and by the MLP model (bottom three rows) evaluated at different thresholds, in the simulated Solar neighbourhoods of galaxies G29, G30, G34 and G42.}
    \label{tab:comp_trad_methods}
\end{table*}

\begin{figure*}
    \centering
    \includegraphics[width=1\textwidth]{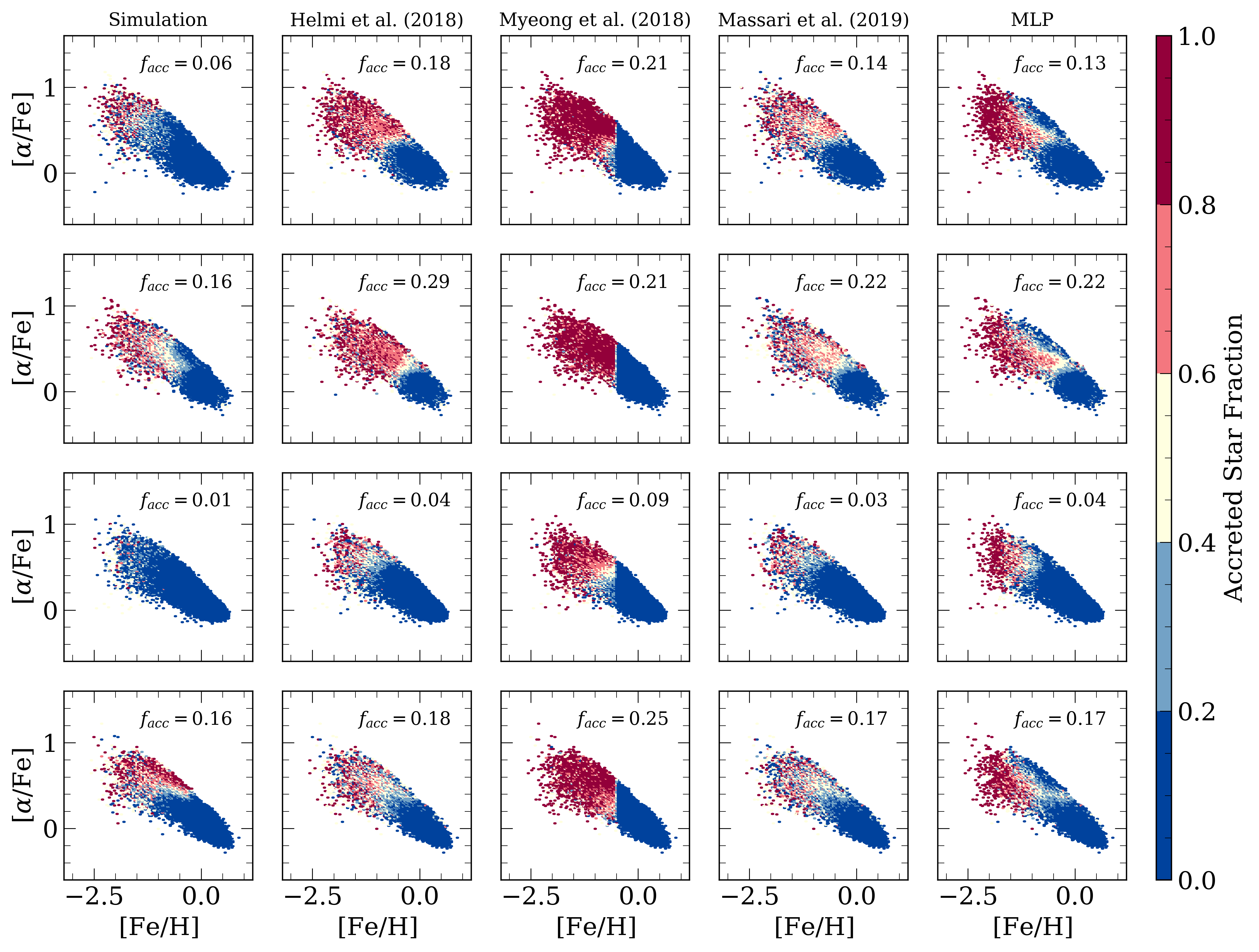}
    \caption{The [$\alpha$/Fe] -- [Fe/H] distribution,  colour-coded by the fraction of accreted stars, $f_{acc}$ for stars in the simulated Solar neighbourhoods. From top to bottom, the rows correspond to galaxies G29, G30, G34, and G42, respectively. In the columns, the accreted stars are defined by: 1) the simulation label; 2-4) the observational selection criteria; 5) the label predicted by the MLP model. In each panel, we show the actual (column 1) and predicted (columns 2-5) overall fractions of accreted stars in the simulated Solar neighbourhoods.}
    \label{fig:comp_trad_method_alpha_iron}
\end{figure*}

\begin{figure*}
    \centering
    \includegraphics[width=1\textwidth]{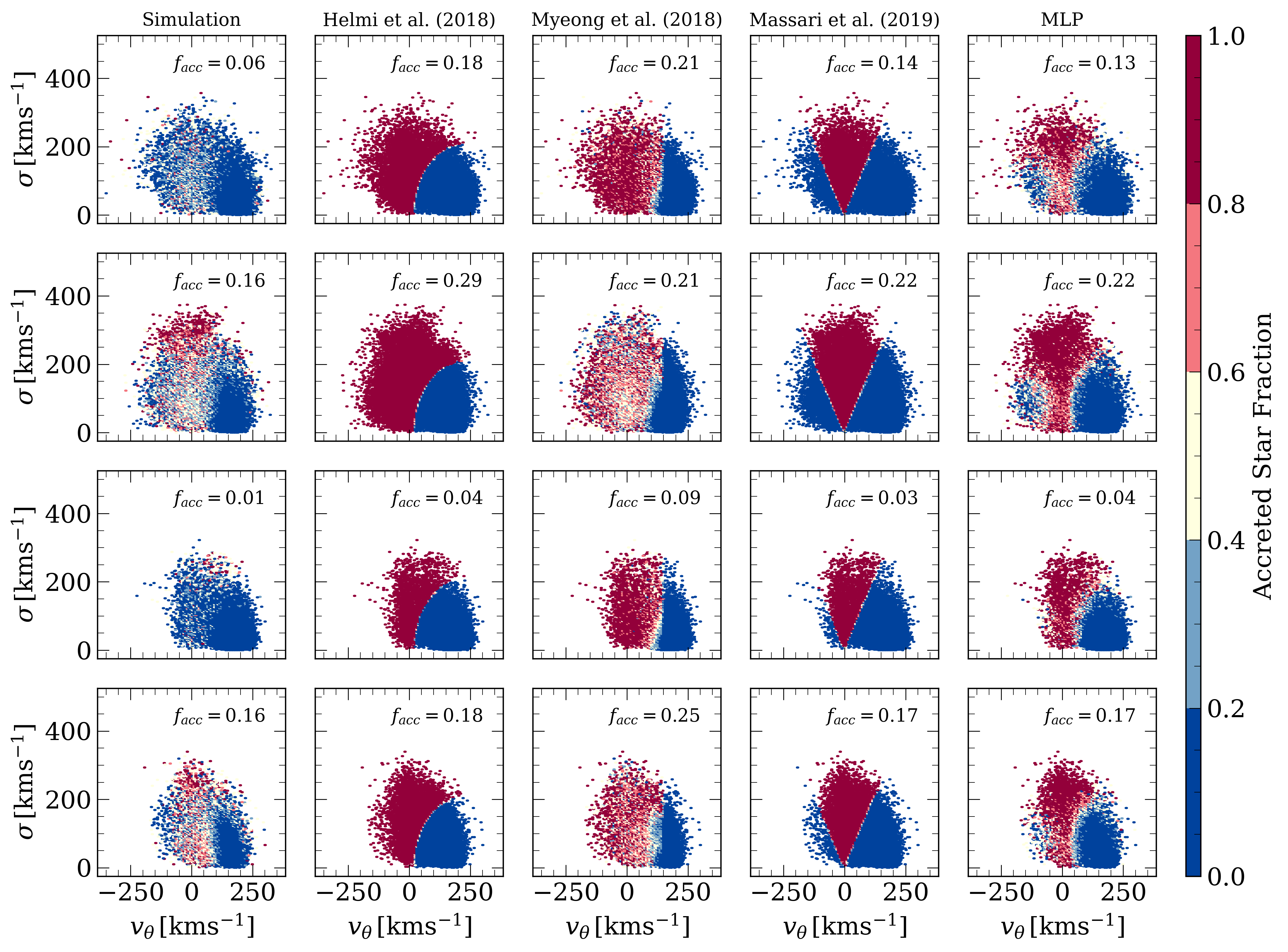}
    \caption{Same as in Fig.~\ref{fig:comp_trad_method_alpha_iron}, but showing results for the distributions of accreted stars in the Toomre diagram.}
    \label{fig:comp_trad_method_toomre}
\end{figure*}

Observational studies of accreted substructures in the Milky Way often focus on regions dominated by halo stars, for example away from the disc. For instance, halo stars are often identified by imposing cuts in the stellar rotational velocities, vertical distances above or below the disc plane, metallicities, or a combination thereof. These cuts are physically motivated, however they may exclude also accreted structures embedded in or near the disc. In the previous sections, we have shown that ML methods are able to identify accreted stars even in the regions which are dominated by in-situ stars (the fractions of FPs and FNs are low even in the disc). This suggests that it may be feasible to apply them directly on the observational data, without making recourse to specific selection cuts. To investigate this possibility, we choose a few representative examples of selection cuts from the literature, and compare the composition of accreted versus in-situ stellar distributions in these cases with the corresponding distributions predicted by the ML models\footnote{We note that, in the analysis of observations, these types selection cuts are just a first step, and further algorithms are applied to the samples to identify accreted substructure/tidal streams, e.g., \texttt{HDBSCAN}, or k-means. As already mentioned, the motivation of this exercise is to investigate whether ML methods could discover more accreted stars in the regions that are typically overlooked by observational methods.}. 
 
Specifically, we consider three examples of selection criteria for halo stars in the Galactic Solar neighbourhood: i) a cut in velocity, $\left| \boldsymbol{V} - \boldsymbol{V_{LSR}} \right| > \mathrm{210} \, \mathrm{km}\,\mathrm{s}^{-1}$ \citep{helmi_merger_2018, koppelman_one_2018, lovdal_substructure_2022}, where the velocity of the Local Standard of Rest is $\boldsymbol{V_{LSR}}=232$ km \,s$^{-1}$ \citep{McMillan2017}; ii) a selection in the $[\mathrm{Fe}/\mathrm{H}]$ - $v_{\theta}$ plane, used by \citet{myeong_halo_2018} to remove the disc stars in order to find accreted substructures in the halo, namely imposing $[\mathrm{Fe}/\mathrm{H}]<-0.5$ and $ v_{\theta}<150 \, \mathrm{km}\,\mathrm{s}^{-1}$; and iii) a kinematic selection used by \citet{massari_origin_2019} to construct a sample of accreted globular clusters, by imposing $\varepsilon>0.5$, where $\varepsilon=L_z/L_{z,circ}$ (the latter was used for larger regions of the Galaxy, however we adapt it here for the Solar neighbourhood). 

For this comparison, we focus on the Solar neighbourhood regions in the simulations, which are defined as ring tori with a minor radii of $2.5$~kpc. The major radii of the tori are determined on a galaxy-by-galaxy basis, by multiplying the Solar radius in the Milky Way (assumed here to be $8$~kpc) with the ratios between the disc-scale lengths of the simulated galaxies and the scale-length of the thin disc of the Milky Way, taken as $2.6$~kpc \citep{Bland-Hawthorn2016}. This accounts for the differences in size between the disc of the Milky Way and the discs of the simulated galaxies. 

We apply the three selection criteria above on the four galaxies from the test dataset (G29, G30, G34 and G42) and label stars as accreted or in-situ according to these cuts. In doing this, we assume that halo stars in the Solar neighbourhoods are the equivalent of stars of accreted origin, and disc stars are equivalent to those formed in-situ. Note that, while these selection criteria are designed to exclude most of the stars of in-situ origin, the real compositions are a mixture of accreted and in-situ, both in and outside the cut-out regions. The fractions of accreted stars in these two regions vary from galaxy-to-galaxy, and also on the type of selection cut that is applied. The fractions of accreted stars in the disc-like regions are very small, although the accreted stars in these regions are likely to be representative of early merger events (i.e., mostly old, metal-poor stars). The labels assigned by these cuts are then compared with the true labels obtained from simulations, and we compute the equivalents of FPs and FNs.

We then apply the ML models on the simulated Solar neighbourhoods in the test galaxies (this time, without any selection cuts) to predict the accreted stars in these regions. As before, we use the models trained on the optimal set of features. For the sake of conciseness, we only present here the results for the MLP, but note that the XGBoost and TML models have similar classification performances. 

In Fig.~\ref{fig:comp_trad_method_alpha_iron} we compare the distributions in the [$\alpha$/Fe] -- [Fe/H] plane of accreted and in-situ stars in the Solar neighbourhood regions predicted by the three selection criteria and by the MLP model. A similar comparison is shown in Fig.~\ref{fig:comp_trad_method_toomre} for the distributions in the Toomre diagram. Both figures illustrate the difference in complexity between the two approaches, with the MLP model being able to provide a closer description of the true distribution of accreted stars in the chemo-dynamical space for all the test galaxies. This result is expected, considering that traditional observational methods are based on the assumption of a simple disc-halo dichotomy. For example, the selection cut of \citet{myeong_halo_2018}, which assumes a distinct dichotomy in terms of [Fe/H], over-predicts the number of accreted stars the [$\alpha$/Fe] -- [Fe/H] space compared with the other methods that do not employ a [Fe/H] cut (see Fig.~\ref{fig:comp_trad_method_alpha_iron}). Overall, the MLP retrieves the accreted stars most accurately, compared with all the selection cut methods. This is the case not only in the overall distribution of accreted stars in the chemical abundance space, but also in terms of the number of accreted stars (compare, for example, the accreted fractions, $f_{acc}$, in the corresponding panels for each galaxy in Fig.~\ref{fig:comp_trad_method_alpha_iron}). 

Likewise, since all selection cuts employ some type of $v_{\theta}$ threshold, they all under-perform compared with the MLP model (see Fig.~\ref{fig:comp_trad_method_toomre}). The criterion of \citet{massari_origin_2019}, which allows for the inclusion of counter-rotating stars in the disc, gives a closer description of the distribution of stars in the Toomre diagram than the other two selection criteria, although it still provides a very simplified version of the true distribution of the two populations. As in the case for chemical abundances, the MLP model is able to retrieve the overall patterns in the kinematical distribution of the two stellar populations, in all four test galaxies. 

Table~\ref{tab:comp_trad_methods} includes the purity ($P$) and completeness ($R$) of the samples of accreted stars retrieved by the MLP model, using three different classification thresholds (the fiducial value of 0.33, 0.10 and 0.75). These metrics are compared with the equivalent P and R values computed using the labels inferred from the three selection cuts versus the true labels from simulations. All values are computed for the Solar neighbourhoods in each of the four galaxies in the test data set. At the fiducial threshold, the MLP model retrieves the purest samples of accreted stars for all test galaxies. When the classification threshold is lowered to 0.10, the model identifies consistently more than 95\% of the accreted stars, while maintaining a precision level very similar to the purest sample retrieved by the selection criteria.  As the classification threshold is increased, progressively purer samples of accreted stars are retrieved, however, at the expense of completeness. When the threshold is set to 0.75, the MLP model is able to create samples of accreted stars, on average, twice as purer as the ones obtained through the observational selection criteria.

In addition to being more accurate in identifying the accreted stars than the selection cuts, the MLP model also retrieves fewer stars labelled accreted (on average, ~20\% fewer than using the selection cuts). This makes it less computationally expensive, especially when applied on large observational datasets. This could be the preferred methodology for the initial processing of observational data to use for subsequent analysis, for example using clustering algorithms to identify tidal stellar streams.

\subsection{Visualisation of accreted and in-situ structures with UMAP}
\label{subsec:umap_results}

\begin{figure*}
    \centering

    \includegraphics[width=1\textwidth]{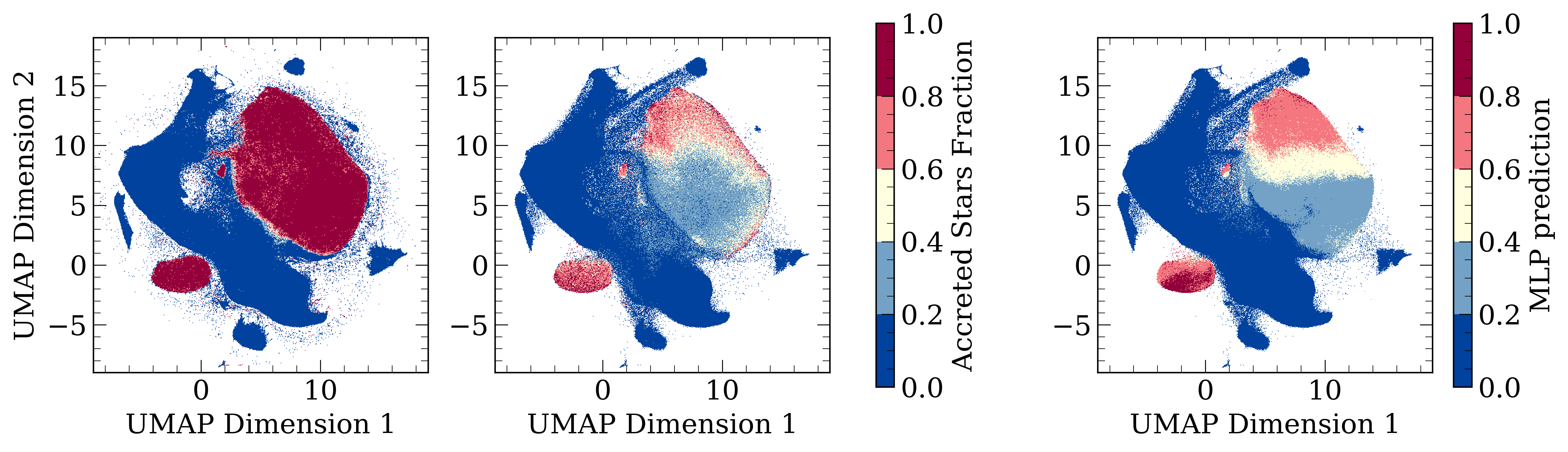}
    
    \caption{Distribution of stars in the training (left panel) and test (central and right panel) datasets in the parameter space defined by UMAP to maximise the separation between the accreted and in-situ populations. Colours represent the fraction of accreted stars in each region of the plane, as defined by the simulation labels (left and central panels) and by the predictions of the MLP model (right panel).}
    
    \label{fig:umap}
\end{figure*}

\begin{figure*}
    \centering
    \includegraphics[width=1\textwidth]{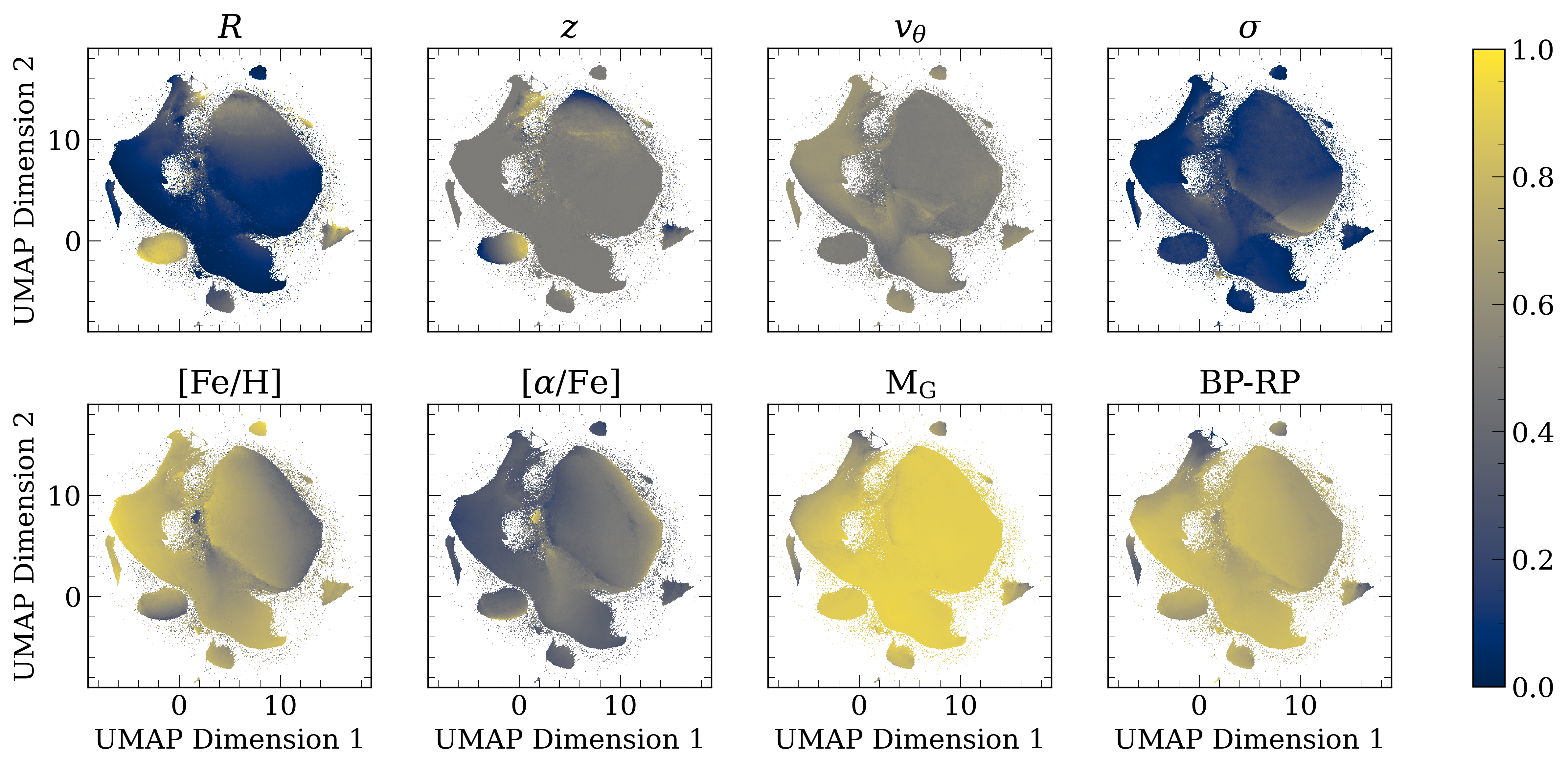}
    \caption{UMAP projections of the distribution of accreted and in-situ examples in the training data, colour-coded by the value of the corresponding physical parameters from the optimal set of features, used as input to the UMAP model. All values are normalised.} 
    \label{fig:umap_train_prop}
\end{figure*}

Further insights on the behaviour of the ML models can be obtained from the distribution of accreted and in-situ stars in a 2D plane constructed by the UMAP model. Through the UMAP algorithm, stars with similar properties are pulled together, revealing structures which may be hidden in the usual spatial or chemo-dynamical parameter spaces. 

In Fig.~\ref{fig:umap} we show the distribution of the training and test examples obtained using the UMAP model. The 2D planes are colour-coded based on the fraction of accreted stars, using the simulation labels (left and middle panels) and the predictions of the MLP model (right panel), respectively. As before, the MLP model is considered representative of all ML models. The training distribution indicates a clear separation between the accreted and in-situ stars. The accreted stars appear to be clustered in two main regions: the smallest cluster comprises stars located outside $R\simeq 15$ kpc, while the other includes stars with a broader range of properties, which are also more gradually-changing. The stars in this second cluster typically have high velocity-dispersions and low [Fe/H] abundances. The in-situ stars appear to be mostly clustered in an L-shaped region characterised by high-rotational motions. Smaller clusters of in-situ stars with specific features are also present in this plane. For instance, the top cluster in this figure is comprised of young, metal-rich stars, located in the inner region of the galaxy; while the right-most cluster is composed of stars orbiting in the plane of the disc, but at large distances from the centre of the galaxy. 

In Fig.~\ref{fig:umap_train_prop} we show the distribution of the training data, colour-coded by each of the stellar properties in the optimal set of input parameters. All properties are presented as normalised values. The same structures are observed in the distribution of the test examples once projected in the UMAP-defined space (central panel, Fig.~\ref{fig:umap}); however, the largest region of the accreted stars shows a significant contamination of in-situ examples. This seems to suggest that some in-situ stars in the test dataset have stellar parameters that resemble those in the accreted examples from the training dataset. As expected, these are the examples which the ML models struggle to associate to either the accreted or the in-situ classes (as shown in the right panel in Fig.~\ref{fig:umap}), and which comprise the majority of the misclassified stars (in both the FP and FN cases). Moreover, the MLP predictions in the largest accreted region appear to follow a gradient as the model outputs progressively higher probabilities of being accreted, for stars occupying the higher parts of this region. In Fig.~\ref{fig:umap_train_prop} a similar, although much shallower, gradient is observed in the $R$-plot, suggesting that the accreted stars from late mergers are identified with higher probabilities by the model, as already concluded from Fig.~\ref{fig:FN}.

\section{Testing the models on the Auriga simulations}
\label{sec:cross_sim}

We further test the performance of our models with an independent dataset, drawn from the \texttt{Auriga} simulations \citep{grand_auriga_2017}. This allows us to investigate whether there are any biases introduced in our models due to training only on the \texttt{ARTEMIS} simulations, and to test the classification performance of the models outside their development environment. For this purpose, we use the ``level 3'' set of six  galaxies from the Auriga Project public data release \citep{grand_data_release_2024}: Au6, Au16, Au21, Au23, Au24 and Au27. These are also disc galaxies of Milky Way mass, with total masses ranging between $1.04 - 1.74 \times 10^{12}\, {\rm M}_{\odot}$ and disc-to-total ratios, $D/T$, ranging from 0.63 to 0.83 \citep{grand_auriga_2017}.

The simulations were run with the hydrodynamical code \texttt{AREPO} \citep{springel_arepo_2010}, which includes physical subgrid prescriptions \citep{vogelsberger_galaxy_2013} that are significantly different from those implemented in the \texttt{EAGLE} code which was used for \texttt{ARTEMIS}.  Furthermore, these simulations have somewhat higher numerical resolution than \texttt{ARTEMIS}, with dark matter particle masses of $\sim4\,\times\,10^{4}\,\textrm{M}_{\odot}$ and baryonic masses of $\sim5\,\times\,10^{3}\,\textrm{M}_{\odot}$, respectively. However, we find that the simulated galaxies in \texttt{Auriga} have similar spatial distributions of in-situ of accreted stars as in the \texttt{ARTEMIS} (not shown here). The level of differences in the spatial distributions between galaxies drawn from these two suites of simulations is comparable with the differences seen between galaxies from the same suite.

We apply the same ML models on these six systems and test their performance. First, we divide the sample into a training and a test dataset, setting aside a fraction of stars in the training dataset for validation purposes. Galaxies Au6 and Au21 experienced the most massive merging events with mass ratios of 0.54 and 0.53 \citep{grand_origin_2018}, respectively, and were thus selected for the test dataset. The rest of the galaxies in the sample (Au16, Au23, Au24, and Au27) were used to provide training examples to the models. 

\begin{figure}
    \centering
    \includegraphics[width=0.9\columnwidth]{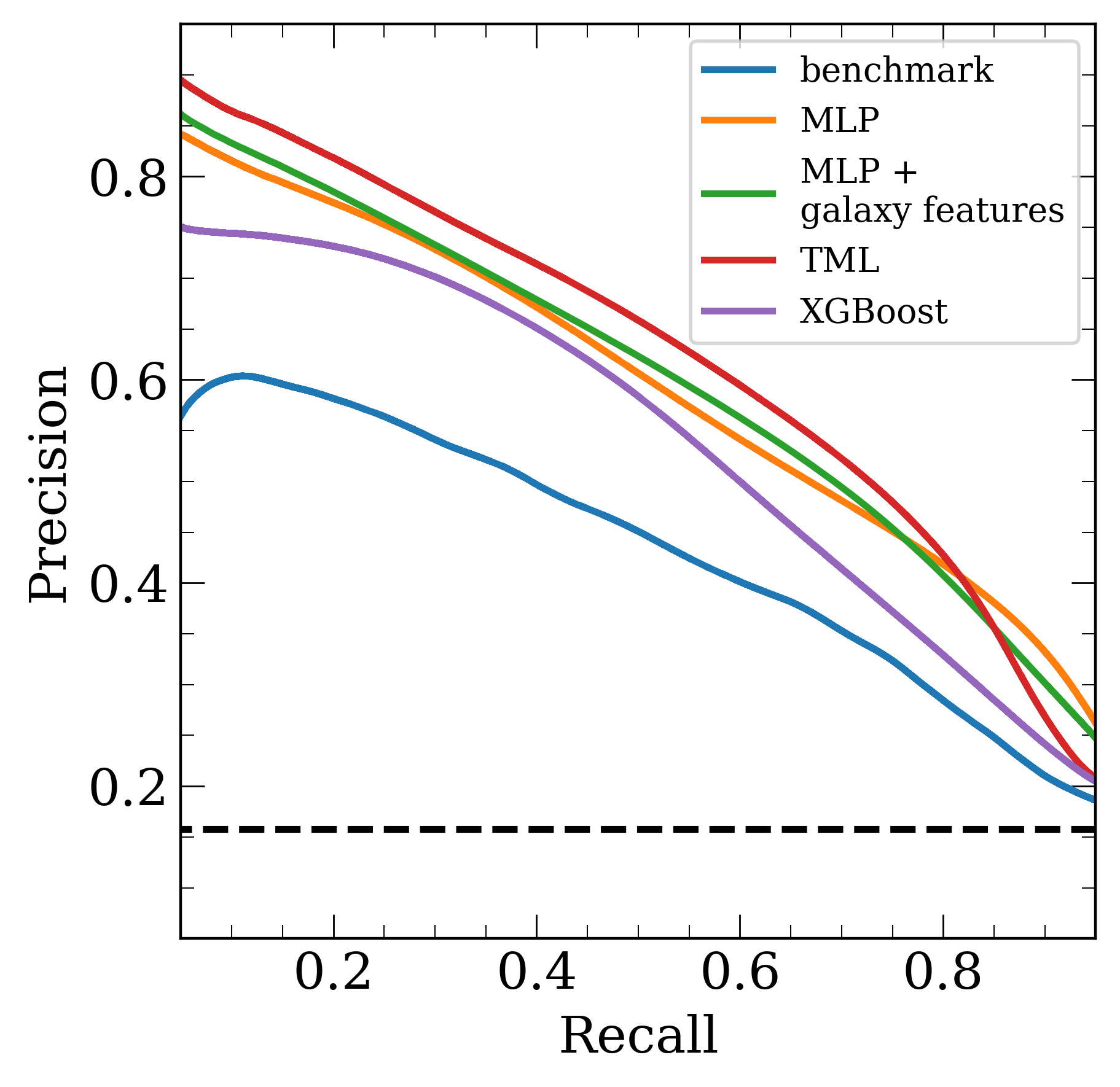}
    \caption{Precision ($P$) and recall ($R$) at different classification thresholds for
the benchmark model, MLP, MLP+ galaxy features, TMP and XGBoost, trained on stars from six \texttt{Auriga} galaxies. The metrics are evaluated considering
all stars in galaxies Au6 and Au21.}
    \label{fig:PR_auriga}
\end{figure}

Fig.~\ref{fig:PR_auriga} shows how the ANNs and decision-tree models perform when developed on the \texttt{Auriga} data. The MLP, MLP+galaxy features and the TML models have a slightly better classification performance than XGBoost. This may be caused by a missed optimisation of the XGBoost hyper-parameters (as we use the the same hyper-parameters tuned for the \texttt{ARTEMIS} data). Alternatively, there could be more complex non-linear relations between the features describing the accreted stars which are better modelled by the ANNs. Nevertheless, similarly to what it was found for \texttt{ARTEMIS}, all these four models have a similar classification performance. This is encouraging, as it suggests that the models are able to extract the relevant relationships between the accreted and in-situ stars, regardless of the type of simulation they were developed on. The benchmark model shows an improved performance compared to the \texttt{ARTEMIS} analogue (PR-AUC score of 0.45 versus 0.36) suggesting a clearer chemical distinction between accreted and in-situ stars in the \texttt{Auriga} galaxies. Nevertheless, the benchmark model returns samples of the lowest purity compared to the other models developed in \texttt{Auriga}, showing that, as for \texttt{ARTEMIS}, the addition of kinematic and photometric information improves significantly the classification.

The MLP, MLP+galaxy features and the TML models share a similar classification performance, with PR-AUC scores of 0.59, 0.60, and 0.62, respectively, while the XGBoost model shows a drop in performance (0.53). These scores are similar to those found for \texttt{ARTEMIS}, which are 0.59 (MLP), 0.55 (MLP+galaxy features), 0.57 (TML), 0.59 (XGBoost), respectively  (see Section~\ref{sec:results}).

\begin{figure}
    \centering
    \includegraphics[width=1.1\columnwidth]{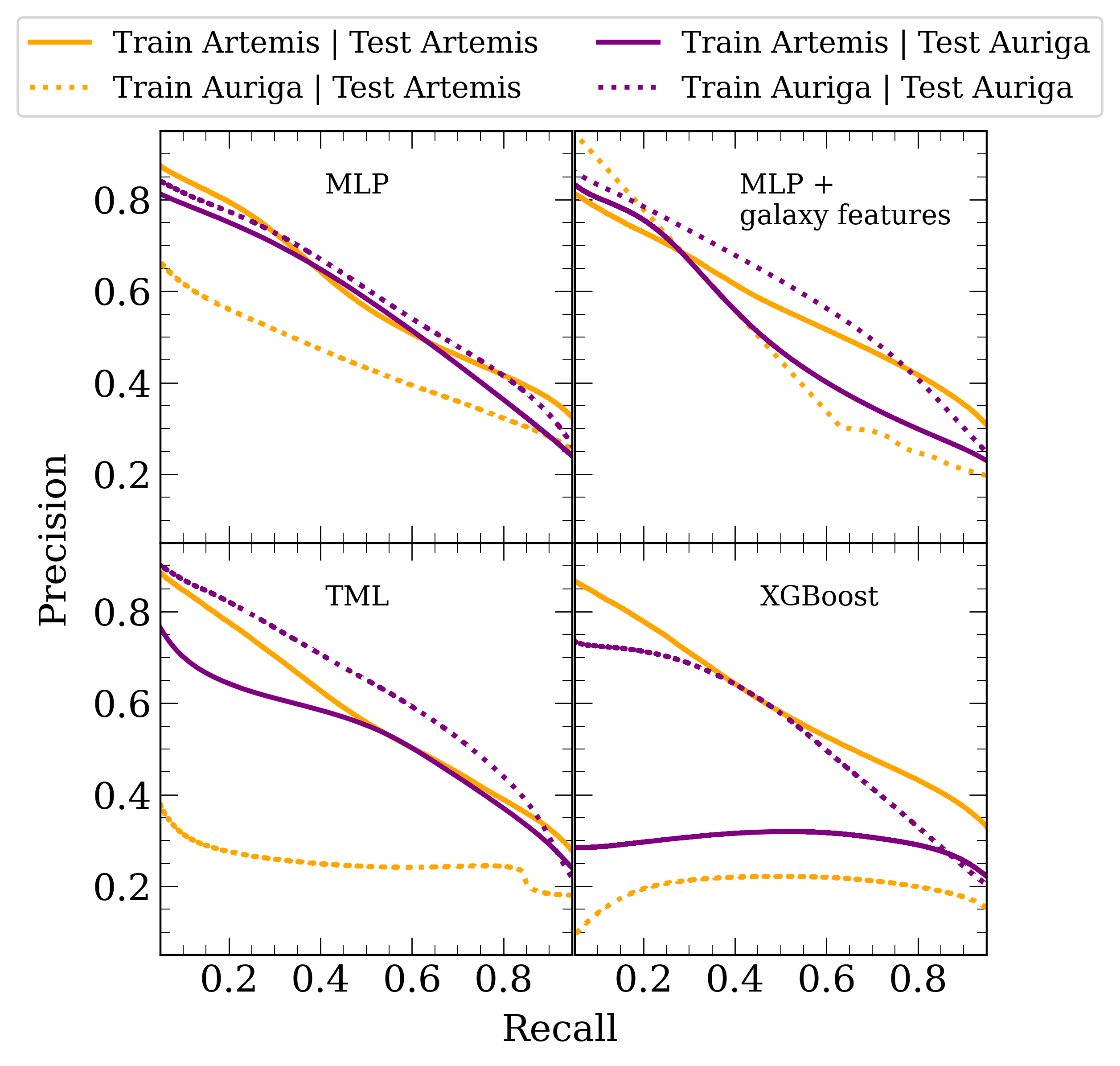}
    \caption{Comparison of the classification performance of  models trained on \texttt{ARTEMIS} and tested on \texttt{Auriga}, and vice versa (orange lines). Included are also the performance of models trained and tested on data from the same simulation (purple lines).}
    \label{fig:cross_sims_comp}
\end{figure}

We perform an additional test, whereby the models developed on the training dataset from \texttt{ARTEMIS} are applied on the \texttt{Auriga} test dataset, and vice versa (purple lines and orange dots in Fig.~\ref{fig:cross_sims_comp}, respectively). This allows us to further investigate the robustness of the models. We also compare the performance of criss-crossing the simulations with the performance of using sets from the same type of simulation (shown by orange lines for \texttt{ARTEMIS} models and purple dots for the \texttt{Auriga} ones in Fig.~\ref{fig:cross_sims_comp}).

Interestingly, when the models trained on \texttt{Auriga} are applied to the test data from \texttt{ARTEMIS}, the classification performance drops drastically. The same is observed when the models trained on the \texttt{ARTEMIS} data are applied to the \texttt{Auriga} test data. The lack of consistency in the classification performance of the model can be explained in terms of both data and model complexity.

Differences in the code, physical model, or the numerical resolution of the two simulations inevitably result in a domain shift between the two datasets; consequently, the models trained and tested on different simulations show a drop in classification performance. Domain adaptation techniques, such as described in \citet{ciprijanovic_deepmerge_2020}, could be explored to develop models that can maintain a consistent classification performance when applied across simulations or on observational data.
Models with a high level of complexity can be affected by overfitting as they capture simulation-specific patterns while learning the distinction between accreted and in-situ stars. Combined with domain shift, overfitting leads to more drastic performance drops. As shown in Fig.~\ref{fig:cross_sims_comp}, this is the case of the XGBoost model, which has sufficient complexity to extract external patterns from the data, being effectively fine-tuned on the set of simulations it is trained on. The MLP with galaxy features model also performs better on the simulation it is trained on. This can be due to the galaxy-specific properties may be affected by the different galaxy formation models used in \texttt{ARTEMIS} and \texttt{Auriga} simulations. For the MLP and TML methods, the models developed on \texttt{ARTEMIS} data have a better out-of-sample classification performance then their \texttt{Auriga} counterparts. Despite being similar in size, the \texttt{ARTEMIS} training dataset comprises a larger number of assembly histories leading to a wider variety of accreted star properties learned by the models.

Fig.~\ref{fig:cross_sims_comp} shows that the MLP model trained on \texttt{ARTEMIS} data is the only model that maintains a consistent classification performance across the two simulations. This is probably due to its simpler model architecture, which makes it less sensitive to overfitting, as well as to the larger sample of assembly histories in the \texttt{ARTEMIS} training set (12 galaxies versus 4 in the \texttt{Auriga} set). Because of its more consistent performance when applied to an entirely different set of simulations, the MLP model is perhaps better suited to be applied on entirely unseen data, such as observational data from the Milky Way.

\section{Conclusions}
\label{sec:conclusions}

In this study we have investigated the performance of different ML models in separating accreted from in-situ stars in Milky Way-mass galaxies, using data from the \texttt{ARTEMIS} simulations. We developed and compared models based on ANN and decision trees algorithms and compared the performance of these models using the usual performance metrics and other physically-motivated diagnostics.  

The main conclusions of this study are as follows:

\begin{itemize}

    \item We find that the optimal set of stellar features for the models includes a combination of positions and kinematics ($R, z, v_{\theta}, \sigma$), photometry ($M_{G}$, $BP-RP$), and chemical abundances ($[\alpha/\mathrm{Fe}]$ and $[\mathrm{Fe}/\mathrm{H}]$). A similar classification performance is found with a slightly reduced set of features, including positions, kinematics and photometry (Fig.~\ref{fig:features_selection}).

    \item All ML models investigated here have good classification performances (Fig.~\ref{fig:PR_test}). Moreover, MLP, TML, and XGBoost perform similarly well also in terms of the distribution of FPs and FNs in a chemo-dynamical parameter space (Figs.~\ref{fig:FP} and \ref{fig:FN}). The majority of mis-classified accreted stars are within a $5 \, \mathrm{kpc}$ radius from the centres of the test galaxies. However, the models perform reasonably well in the regions dominated by in-situ stars (e.g., the disc), and typically, the fractions of FPs and FNs are only a few percent. In the outer regions of galaxies, the MLP, TML, and XGBoost models retrieve more than 90\% of the accreted stars with an accuracy close to 60\%, at the fiducial thresholds. These results are remarkable, given that these models belong to different families of ML methods, suggesting that the similar underlying patterns related to accreted and in-situ stars are learned in all cases. Similar patterns are also retrieved by the UMAP dimensionality reduction method (Figs.~\ref{fig:umap} and \ref{fig:umap_train_prop}).
    
    \item Of all models investigated here,  MLP is less sensitive to performance drops due to domain shift related to the specificity of accretion history of any given galaxy, and could be the preferred option to use on observational data. However, developing an ML model that is able to learn the galaxy-specific properties (namely, the accretion histories of galaxies) remains a challenge. We have found that adding average galaxy properties, such as stellar masses or metallicities, to the set of input features does not improve the classification performance (specifically, the MLP+galaxy features model), and on the contrary, it leads to overfitting (Fig.~\ref{fig:PR_test}).  Also, combining the predictions of multiple models using ensemble learning (i.e., the TML model) does not provide an improvement, and more complex domain adaptation techniques may be needed to address this issue.

    \item As shown by the importance gain in the XGBoost model, the parameters which provide the most accurate distinction between the two populations are: the rotational velocity ($v_{\theta}$), the galactocentric distance in the plane of the disc ($R$), and the $[\mathrm{Fe}/\mathrm{H}]$ abundance (Fig.~\ref{fig:xgb_features_importance}).

    \item The purity ($P$) of the sample of accreted stars retrieved by the models can be increased by adjusting the classification threshold (Fig.~\ref{fig:diff_thresh_energy}), however this comes at the cost of completeness ($R$). The accreted stars identified more accurately by the models have chemo-dynamical properties associated with late accretions, and are located in the outer regions of galaxies. This suggests that adjusting the classification threshold of ML models can also be used to identify tidal streams in the outer halo.  

    \item We also compared the performance of the ML models versus  imposing common observational selection cuts (either in space, kinematics or chemistry) to separate accreted stars from those formed in-situ. We have found that ML models outperform in purity these more traditional methods (Figs.~\ref{fig:comp_trad_method_alpha_iron} and \ref{fig:comp_trad_method_toomre}). Therefore, ML models may be applied directly on observational data without the need of additional selection criteria. Thus, they may help in the search for accreted substructures  even in the regions dominated by the disc. 

    \item  Finally, we have tested the models on a different suite of cosmological simulations (namely, on \texttt{Auriga}), to evaluate their performance on unseen data (Figs.~\ref{fig:PR_auriga} and ~\ref{fig:cross_sims_comp}). In general, we find that the models perform similarly well on \texttt{Auriga} as on \texttt{ARTEMIS}, which suggests that they may be also suitable to be applied on other types of previously unseen data, for example, on observations. Of all the models, XGBoost has the least performance on an unseen dataset, possibly because it uses more detailed properties, which differ between the two sets of simulations. In contrast, the MLP appears to be using more broadbrush properties that are relevant to the overall trends between features.  These results highlight the importance of testing not only of different ML models, but of different training sets as well.
    
\end{itemize}

Our study has shown that ML methods can efficiently separate accreted from in-situ stars in galaxies like the Milky Way. These methods perform optimally with a combination of kinematics and chemical abundances, and can improve the detection of accreted substructures in regions of the Galaxy that are highly dominated by in-situ stars, and which have not been fully explored to date for identifying substructure. A wealth of high precision data are already available for millions of Milky Way stars, from both astrometric observations, e.g. with {\it Gaia}  \citep{gaia_collaboration_gaia_2018,Gaia2023}, and spectroscopic measurements of chemical abundances, from surveys such as APOGEE \citep{majewski_apache_2017}, GALAH \citep{de_silva_galah_2015}, LAMOST \citep{zhao_lamost_2012},  WEAVE \citep{dalton_project_2014} or 4MOST \citep{de_jong_4most_2019}. ML methods such as the ones developed here can be directly deployed on this combined, multi-dimensional parameter space to help in the discovery of accreted substructures.

\section*{Acknowledgements}

AS acknowledges an STFC PhD studentship at the LIV.INNO Centre for Doctoral Training ``Innovation in Data Intensive Science''. The models have been trained and tested on LJMU's HPC facility, Prospero.
IGM has received funding from the European Research Council (ERC) under the European Union's Horizon 2020 research and innovation programme (grant agreement No 769130).  The construction of the \texttt{ARTEMIS} simulations used the DiRAC@Durham facility managed by the Institute for Computational Cosmology on behalf of the STFC DiRAC HPC Facility. The equipment was funded by BEIS capital funding via STFC capital grants ST/P002293/1, ST/R002371/1 and ST/S002502/1, Durham University and STFC operations grant ST/R000832/1. DiRAC is part of the National e-Infrastructure.
\section*{Data Availability}

 All codes and models are publicly available at \url{https://github.com/ariasant/ML-accreted-vs-insitu}. Data from the \textsc{artemis} simulations may be shared on reasonable request to the corresponding author.



\bibliographystyle{mnras}
\bibliography{refs} 





\bsp	
\label{lastpage}
\end{document}